\documentclass{aastex631}

\usepackage{ragged2e}
\usepackage{wrapfig}
\usepackage{float}
\usepackage{multirow}
\usepackage[version=4]{mhchem}
\usepackage{tablefootnote}

\definecolor{EDIT}{rgb}{0.0, 0.5, 0.25}

\begin{document}



\title{A Disintegrating Rocky World Shrouded in Dust and Gas: Mid-IR Observations of K2-22b using JWST}


\definecolor{NTT}{rgb}{0.0, 0.5, 0.5}
\newcommand{\NTT}[1]{\textcolor{NTT}{\bf [NTT: #1]}}

\newcommand{\PSUAA}{Department of Astronomy and Astrophysics, The Pennsylvania State University, University Park, PA, 16802, USA}
\newcommand{\PSUGS}{Department of Geosciences, The Pennsylvania State University, University Park, PA, 16802, USA}
\newcommand{\CEHW}{Center for Exoplanets and Habitable Worlds, The Pennsylvania State University, University Park, PA, 16802, USA}
\newcommand{\Wisc}{Department of Astronomy, University of Wisconsin-Madison, Madison, WI, USA}
\newcommand{\NASA}{NASA Goddard Space Flight Center, Greenbelt, MD, 20771, USA}
\newcommand{\ASU}{School of Earth and Space Exploration, Arizona State University, PO Box 876004, Tempe, 85287-6004, Arizona, USA}
\newcommand{\IMP}{Astrophysics Group, Imperial College London, Blackett Laboratory, Prince Consort Road, London SW7 2AZ, UK}
\newcommand{\Ox}{Sub-department of Astrophysics, Department of Physics, University of Oxford, Keble Road, Oxford, OX1 3RH, UK}
\newcommand{\MaxP}{Max-Planck-Institut für Astronomie, Königstuhl 17, D-69117 Heidelberg, Germany}
\newcommand{\Bern}{University of Bern}
\newcommand{\Dminn}{Department of Physics and Astronomy, University of Minnesota Duluth, Duluth, Minnesota 55812, USA}
\newcommand{\NOIR}{U.S. National Science Foundation National Optical-Infrared Astronomy Research Laboratory, 950 N. Cherry Ave., Tucson, AZ 85719, USA}

\correspondingauthor{Nick Tusay}
\email{tusay@psu.edu}

\author[0000-0001-9686-5890]{Nick Tusay}
\affiliation{\PSUAA}
\affiliation{\CEHW}

\author[0000-0001-6160-5888]{Jason T. Wright}
\affiliation{\PSUAA}
\affiliation{\CEHW}

\author[0000-0002-9539-4203]{Thomas G. Beatty}
\affiliation{\Wisc}

\author[0000-0002-1571-0836]{Steve Desch}
\affiliation{\ASU}

\author[0000-0001-8020-7121]{Knicole Col\'on}
\affiliation{\NASA}

\author[0000-0002-8026-0018]{Tushar Mittal}
\affiliation{\PSUGS}

\author[0000-0002-4047-4724]{Hugh~P.~Osborn}
\affiliation{NCCR/Planet-S, Physikalisches Institut, Universität Bern, Gesellschaftsstrasse 6, 3012 Bern, Switzerland}
\affiliation{Inst. f. Teilchen- und Astrophysik, ETH Zürich, Wolfgang-Pauli-Strasse 27, 8093 Zürich, Switzerland}

\author[0000-0003-0225-1201]{Beatriz Campos Estrada}
\affiliation{\MaxP}

\author[0000-0002-4856-7837]{James E. Owen}
\affiliation{\IMP}

\author[0000-0002-2990-7613]{Jessica Libby-Roberts}
\affiliation{\PSUAA}
\affiliation{\CEHW}

\author[0000-0002-5463-9980]{Arvind F. Gupta}
\affiliation{\NOIR}

\author[0000-0003-4543-0648]{Brad Foley}
\affiliation{\PSUGS}

\author[0000-0002-2160-8782]{Erik Meier Vald\'es}
\affiliation{\Ox}

\author[0000-0002-5951-8328]{Daniel J. Stevens}
\affiliation{\Dminn}

\author[0000-0002-6051-9002]{Ashley Herbst}
\affiliation{\ASU}

\begin{abstract}

The disintegrating ultra-short period rocky exoplanet K2-22b periodically emits dusty clouds in a dynamically chaotic process resulting in a variable transit depth from 0--1.3\%. The effluents that sublimate off the surface and condense out in space are probably representative of the formerly interior layers convectively transported to the molten surface. Transmission spectroscopy of these transiting clouds reveal spectral fingerprints of the interior composition of this rocky world. We used JWST’s Mid-Infrared Instrument (MIRI) as a low-resolution slitless spectrograph to observe four predicted transit windows for K2-22b. For each observation, we extracted a transmission spectrum over the spectral range of 4.4--11.8\,$\mu$m. Over the spectral range of 4.4--8\,$\mu$m, where the spectral precision is highest, we detect one transit at high significance and two at low significance. While the S/N of the spectrum limits our ability to draw firm conclusions, we find that the data: 1) disfavor featureless, iron-dominated core material, 2) are consistent with some form of magnesium silicate minerals, likely from mantle material, and 3) show a distinct and unexpected feature at $\sim$5\,$\mu$m. The unexpected feature, also seen weakly in the low-significance transits, is consistent with an unknown gaseous absorber, possibly NO and/or CO$_2$. These findings warrant further study to improve the constraints on the composition of this disintegrating rocky world.

\end{abstract}


\section{Introduction} \label{sec:intro}

Protoplanetary disks \citep[e.g.][]{Thiabaud:2015:}, debris disks \citep[e.g.][]{Hughes:2018:}, and polluted white dwarfs \citep[WD; e.g.][]{Jura2014, Vanderburg2015} provide glimpses into the constituents of planetary bodies from the beginning and end stages of a system's life-cycle. Studies of such systems have also been used to assess the relationships between host star metallicity and the material available in their planets \citep[e.g.]{Wang:2019:, Adibekyan:2021:, Swain:2024:}. However, these approaches, while valuable, are only indirect probes of a planet's composition. Various planetary accretion and geophysical processes can significantly impact the composition and evolution of rocky planets over their star's main sequence phase \citep{Ballmer:2021:, Foley:2024:}. 

The interior composition of terrestrial planets and its coupling with the surface-atmosphere is of particular interest to the astrobiological community for assessing planetary habitability. Interpretation of planet bulk density measurements has severe compositional degeneracies, especially with regard to a planet's interior versus surface volatile reservoirs. These reservoirs are strongly influenced by geophysical processes and, in turn, impact continued planetary habitability \citep{Seager:2007:, Dorn:2015:, Cockell:2024:}. With only the planets in our Solar System for comparison, and no clear understanding of how unique our system is, we need more observations of other systems to tease out details on interior and geophysical processes for extrasolar planets. 
But fully formed planets in stable orbits around main sequence stars have a habit of keeping their rocky interiors on the inside, making the truth of their constituents difficult to access.    
Disintegrating exoplanets offer a unique opportunity to directly probe the interior composition of rocky exoplanets \citep{bodman2018inferring}. As a unique sub-class of ultra-short period planets (USPs, ${\scriptstyle \lesssim}$24hr orbits), disintegrating planets are not close enough to their host star to be gravitationally disrupted but hot enough that their interior layers likely convect with a molten surface that ultimately sublimates out into space \citep{Rappaport_2012}. 

K2-22b is one of only three actively disintegrating rocky USPs detected by the \textit{Kepler} and \textit{K2} missions. K2-22b transits a relatively bright, early M-dwarf ($K=11.91$, $T_{\rm eff}=3830$ K) with an orbital period of 9.1457 hours \citep{sanchis2015k2}. The dramatic variability in lightcurve transit depth (0--1.3\%) combined with the asymmetric transit shape suggests we are observing a transient cloud of dust sublimating off the surface of an otherwise unseen planet \citep{Rappaport_2012}.

Having likely persisted in this state of constantly vaporizing surface material, the deeper layers of the planet may at this point be convecting to the surface or completely exposed through significant mass loss \citep{Perez-Becker_2013PhDT.......300P,Curry1_interior_model:2024:}. Thus, measuring the composition of these dusty effluents can probe the bulk composition of what used to be the interior of the planet \citep{Curry2:2024:}. Previous studies of Kepler-1520b (KIC 12557548b; hereafter “KIC 1255b”) \citep{Rappaport_2012} and KOI-2700b \citep{Rappaport:2014:}, the two other previously known disintegrating USP planets, have shown through dynamical-modeling constraints that the dust in their tails might be composed of corundum (Al$_2$O$_3$ [s]) or iron-rich silicates, while other compositions such as pure iron or graphite have been ruled out \citep{Lieshout_2014, Lieshout_2016}. Subsequent modeling efforts by \citet{Bromley:2023:} and \citet{CamposEstrada:2024:} have found iron-bearing silicates to be the most likely composition of the dust for K2-22b and KIC 1255b. 

In this paper, we present the results of the first mid-infrared spectroscopic measurements of a disintegrating exoplanet, K2-22b. These measurements were made using the Mid-Infrared Instrument (MIRI) in low-resolution spectroscopy (LRS) mode on JWST. 
The primary objective of analyzing these observations is to match any spectral features with opacities from plausible compositions and assess if we are observing minerals and/or gases condensing from vaporization of planetary core or mantle/crust materials.
In Section~\ref{sec:obs}, we describe the JWST observations and simultaneous optical observations with the CHaracterising ExOPlanet Satellite (CHEOPS). In Section~\ref{sec:results}, we present the results of our analysis and the opacity models we used for comparison with the features seen in our spectra. In Section~\ref{sec:disc}, we discuss our interpretation of the results, the implication they have for the composition of this target, the volatility of disintegrating exoplanet activity levels, and future work. 
The appendices include additional information to support our analysis, starting with MIRI data reduction using 
two independent pipelines, \S\,\ref{sec:eureka} \& \S\,\ref{sec:pegasus}, and CHEOPS data reduction as well, \S\,\ref{sec:CHEOPS_reduction}. Then, we have included additional discussion and plots detailing outlier removal, \S\,\ref{sec:app_outliers}, lightcurve fitting and uncertainty assessment, \S\,\ref{sec:app_lightcurves}, supporting evidence for spectral features, \S\,\ref{sec:app_4spectra}, and optical data references, \S\,\ref{sec:app_opt_data}.


\section{Observations and Analysis} \label{sec:obs}
    \subsection{Infrared Observations with JWST}

    \renewcommand{\columnsep}{7pt}
    \begin{wraptable}{r}{0.75\textwidth}
        \vspace{-0.6cm}
        \centering
        \caption{JWST Cycle 2 GO Program 3315 \& CHEOPS DDT-0020 Observation Times}
        \label{tab:obs}
        \vspace{-0.5cm}
        \begin{tabular}{|c|c|c|}
            \hline
            Obs. \# & JWST Start \& End Times (UTC) & CHEOPS Start \& End Times (UTC) \\
            \hline
            1 & April 24, 2024, 16:14:02--19:13:21 & April 24, 2024, 12:24:19--21:54:33 \\
            2 & April 26, 2024 13:57:53--16:57:11 & April 26, 2024, 09:57:08--20:02:22 \\
            3 & April 27, 2024, 08:15:54--14:19:18 & April 27, 2024, 04:34:03--14:10:16 \\
            4 & April 27, 2024, 14:21:05--20:24:29 & April 27, 2024, 14:10:16--23:46:30 \\
            \hline
        \end{tabular}
    \end{wraptable}
        Through GO Program 3315 (PI: Wright, Science PI: Tusay), we used the MIRI LRS instrument in slitless mode on JWST to observe: (a) 2 individual transits of K2-22b, and (b) a full phase curve including 2 additional transits, using the latest available ephemeris at the time of observation scheduling from \citet{Schlawin:2021:}. 
        Table \ref{tab:obs} provides a summary list of the observation times. The data was recorded at a cadence of 72 seconds. All the JWST data used in this paper can be found in MAST: \dataset[10.17909/sxct-na33]{http://dx.doi.org/10.17909/sxct-na33}.

        \renewcommand{\columnsep}{7pt}
        \begin{wrapfigure}{r}{0.4\textwidth}
            \vspace{-0.5cm}
            {\includegraphics[width=0.4\textwidth]{MIRI_ver_img_DoD.png}}
            \caption{MIRI verification image of K2-22 and its smaller companion star, at $\sim$2$^{\prime\prime}$ separation. The direction of dispersion (DoD) is included as well, showing enough separation between the binary that contamination is minimal. The DoD is $\sim$175$^{\circ}$ East of North.}
            \label{fig:ver_img}
            \vspace{-1.0cm}
        \end{wrapfigure}        
        The roll angle of the telescope was such that the dispersion of the light was not contaminated by the companion star, spatially separated by about 2$''$.
        Figure \ref{fig:ver_img} shows the MIRI verification image of K2-22b and its likely bound and much fainter companion ($K=13.93$). The spatial separation is sufficiently wide to create an aperture around the target star that can collect a majority of its light without contamination by the companion. 
        More details on the data reduction and apertures used can be found in Appendix \ref{sec:reduction}. The response of the slitless mode of the spectrograph extends bluer than the slit mode, yielding measurements from 4.4 to 11.8\,$\mu$m.

    \subsection{Optical Observations with CHEOPS}

        CHEOPS is a 30-cm aperture space telescope optimized to observe optical (330 -- 1100 nm) transits of exoplanets \citep{benz2021cheops}. It is located in a nadir-locked low-Earth orbit, meaning that the field rotates once every $\sim$100 minutes, with typically $\sim$40 minutes affected by Earth occultation. 
        
        We obtained three separate CHEOPS observations in a 3-day window via a Director's Discretionary Time (DDT) proposal (DDT-0020, PI: Tusay) in order to constrain the optical depth of K2-22b. These observations were simultaneous with the observations from JWST. We targeted the same individual transits and full phase curve.
        For the individual transits, assuming a 0.8-hr transit window, we allowed for 4.1 hours of baseline before and after each transit. Similarly, we obtained a baseline of 4.1 hours before and after the transits for the full phase curve. See Table \ref{tab:obs} for observation times and Appendix \ref{sec:CHEOPS_reduction} for additional details on the CHEOPS data reduction. Analysis of the resulting lightcurve can be found in Appendix \ref{sec:app_lightcurves}. 


    \section{Results} 
    \label{sec:results}

    \renewcommand{\columnsep}{7pt}
    \begin{wraptable}{r}{0.63\textwidth}
        \centering
        \vspace{-0.5cm}
        \begin{tabular}{c|c|c|c}
            \hline
            \multirow{2}{*}{Obs. \#} & Predicted Transit & Transit Depth [ppm] & Depth [ppm] \\
            & Center (UTC) & (4.4--11.8\,$\mu$m) & (4.4--8\,$\mu$m) \\
            \hline
            1 & April 24, 2024, 18:01:41 & 22 $\pm$ 975 & 52 $\pm$ 253 \\
            2 & April 26, 2024, 15:45:26 & 972 $\pm$ 1335 & 727 $\pm$ 247 \\
            3 & April 27, 2024, 10:02:57 & 388 $\pm$ 545 & 446 $\pm$ 174 \\
            4 & April 27, 2024, 19:11:42 & 1506 $\pm$ 540 & 1627 $\pm$ 168 \\
            \hline
        \end{tabular}
        \caption{The calculated transit ephemeris, using \citet{Schlawin:2021:}, and transit depth measurements from the total combined-light MIRI data for each observation of K2-22b, using Equations \ref{eq:depth} \& \ref{eq:unc} over the full wavelength range of the detector (4.4--11.8\,$\mu$m) and in channels $<$8\,$\mu$m. The full wavelength range shows unreasonably large uncertainties, while the restricted wavelength range shows a transit with high significance (9.7-$\sigma$) in the fourth observation and detections of low significance ($<$3-$\sigma$) in observations 2 and 3.}
        \label{tab:results}
        \vspace{-0.6cm}
    \end{wraptable}

    We report our best fit transit depths in Table \ref{tab:results}, including a detection of the transit of K2-22b at high significance (9.7-$\sigma$) during the 4th expected transit window in the combined (white) light of the MIRI instrument data when combining only channels $<8\mu$m (i.e. where our spectra have the highest signal to noise ratios). We find detections of modest significance in two out of the other three transit windows, as well. The larger uncertainties in the longer wavelength data (8--11.8\,$\mu$m) were deemed too unreliable for accurate measurements. We include them in our figures for completeness, but our analyses largely focused on the spectral channels from 4.4--8\,$\mu$m. The depth uncertainties were calculated assuming uncorrelated, Gaussian noise, which we find reasonable over this wavelength range; an analysis accounting for correlated noise and non-Gaussianity (see Appendix \ref{sec:app_lightcurves}) suggests red noise has only a marginal effect on (${\scriptstyle \lesssim}$1.1$\times$) the uncertainties.
    
    \begin{figure}[htb!]
    \centering
    \includegraphics[width=0.95\linewidth,clip]{T4_spectrum_gasNdust_lowres.png}
    \centering
    \caption{\justifying Results from JWST Cycle 2 GO Program 3315.
    \\\hspace{\textwidth}
    \textbf{\emph{Top}}: Lightcurve of the second half of the phase curve, collected on April 28, 2024, showing a clear transit at the predicted ephemeris. The unbinned time series data (at a cadence of 72 seconds) is plotted in grey in the background. The blue points are binned to a time resolution of 8 minutes. The average 46-minute duration drawn from \citet{sanchis2015k2} is shown as vertical dashed grey lines for reference. Since the transit duration incorporating the cloud of material is not precisely known and potentially variable with wavelength, the baseline flux was calculated from data outside the baseline boundaries set at $1.5\times$ the average 46-minute duration, centered at the ephemeris and shown as vertical dashed green lines. The red line is a 2nd order polynomial model fit to the data during transit and set to zero elsewhere. The residuals of the lightcurve model are shown in the bottom panel.
    \\\hspace{\textwidth} \\
    \textbf{\emph{Middle}}: JWST MIRI spectrum showing a clear detection of a transit from 4.4--8\,$\mu$m \emph{except} at 4.8\,$\mu$m. A flat spectrum is disfavored (reduced $\chi_\nu^2 = 3.56$).
    The unbinned JWST data for each channel are plotted in grey in the background. The binned JWST spectrum in the foreground has a constant spectral resolution of R=17. 
    The horizontal red dashed line shows the overall transit depth from the MIRI lightcurve in the top panel. The dotted curves show the spectra models of plausible gas species from the DACE database \citep{Grimm_2021}. The dashed curves show representative combinations of iron-magnesium-silicate species predicted by \citet{bodman2018inferring,CamposEstrada:2024:}. The solid purple line shows a total opacity model representing a combination of these minerals and gases with features that plausibly correspond to features in the data. 
    CHEOPS photometry is also shown over its 0.33--1.1\,$\mu$m bandpass. The measured CHEOPS transit depth is consistent with the overall transit depth from the MIRI lightcurve.
    \\\hspace{\textwidth} \\
    \textbf{\emph{Bottom Insets}}: The lightcurve and transit models in two channels that illustrate an unexpected spectrum feature with the most significance at $\sim$5\,$\mu$m. The unexpected feature at 5.1\,$\mu$m, consistent with nitric oxide (NO) gas is clearly detected, in contrast to a clear non-detection at 4.8\,$\mu$m (see Figure \ref{fig:Gases}).}
    \label{spectrum}
    \end{figure}
    
    Figure \ref{spectrum} shows the white light time series of the second half of the phase curve, including the only significant transit event detected by JWST and the resulting transmission spectrum. Our statistical analysis of the spectrum does not include the CHEOPS data point. The CHEOPS data was primarily used to confirm no unexpectedly large optical depths during this transit, and we show it only for reference. Assuming uncorrelated white noise and examining the spectrum produced during the 4th transit window, a flat spectrum is disfavored; reduced $\chi_\nu^2 = 3.19$, increasing to $\chi_\nu^2 = 3.56$ when considering only shorter wavelengths ($<8\mu$m), and $\chi_\nu^2 = 8.03$ for the most prominent features seen around 4.4--5.3\,$\mu$m. 
    These reduced $\chi_\nu^2$ values are calculated between the spectrum (R=17) data and the weighted mean of the spectrum (i.e., a flat spectrum).
    
    Apparent spectral features, particularly at shorter wavelengths ($<8\mu$m), are consistent with some combination of expected iron-magnesium-silicate species \citep{bodman2018inferring, CamposEstrada:2024:}, and we find multiple species more favorably fit the data than a flat line (Figure \ref{fig:BIC} shows a $\Delta$BIC comparison). However, the large uncertainties from the single spectrum , especially at the longer wavelengths, mean we cannot definitively distinguish between different peridotite silicates. Still, our statistical tests indicate that the iron-bearing, core-like species were more heavily disfavored than the mantle-like species. Furthermore, after analyzing a large suite of minerals, we do not find a suitable mineral with solid-state features consistent with the 4.4--5.3\,$\mu$m features (and with the spectra at other wavelengths). The transit is most significantly detected at 5.1\,$\mu$m, which we can match with with nitric oxide (NO) gas (see Figure \ref{fig:Gases}), in contrast to a clear non-detection at 4.8\,$\mu$m.

    We note that the MIRI instrument has spectral foldover at the shortest wavelengths \citep{Bouchet:2022:}. While a more rigorous quantitative analysis should take this into account, the potential contamination from this effect is unlikely to entirely account for the short-wavelength features. See Appendix \ref{sec:app_4spectra} for more discussion on this.

    We also detect transits at low significance in two of the other three transit windows (see Figures \ref{fig:all4LCs} \& \ref{fig:4spectra}). 
    Intriguingly, in these two other transit windows, the most significant detections are in the 4.5 and 5.1\,$\mu$m channels, which provides an additional line of evidence that these features in transit window 4 are real absorption features and not a statistical fluke or result of systematic uncertainties.

        \begin{figure}[htb!]
        \centering
            \includegraphics[width=0.92\textwidth]{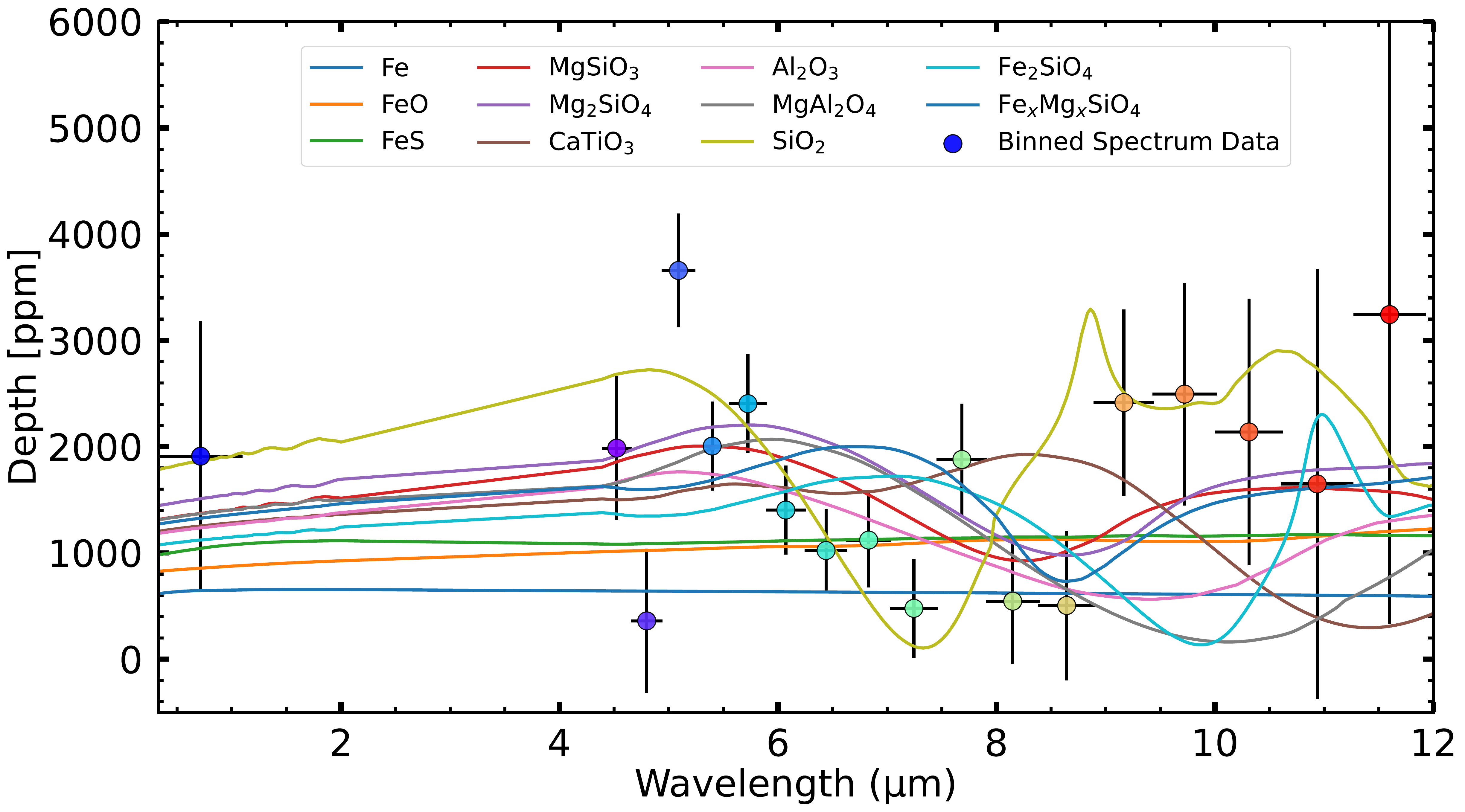}
            \caption{Opacities calculated from optical properties of solid dust species using Mie scattering theory with an effective dust grain size of 1\,$\mu$m for a power law (p = 0.875) distribution of dust grains. Representative models for core, mantle and crust mineralogy are included. This spectrum disfavors purely featureless iron compounds indicative of a bare core. None of these solid dust opacities simultaneously account for all the features in the spectrum, particularly falling short of describing the feature around 5\,$\mu$m.
            \label{fig:solids}}
        \end{figure}
    

\subsection{Solid Dust Features} \label{sec:solids}

We use Mie theory \citep{Mie1908} to compute the opacity of common dust species expected to be abundant in different layers of an Earth-like rocky planet. We pick Earth-like species following previous modeling results which find them to be the most likely \citep{Bromley:2023:, CamposEstrada:2024:}. We assume the dust grain sizes to be power-law distributed between 0.01 and 5\,$\mu$m, with an effective size of 1\,$\mu$m and an index of 0.875. The choice of dust grain sizes is based on morphological modeling of these objects by \citet{sanchis2015k2}, \citet{Lieshout_2016} and \citet{Schlawin:2021:}. 

We considered Fe\,[s], FeO\,[s], and FeS\,[s] as minerals likely to condense after vaporization of planetary core materials. For the mantle, we considered enstatite (MgSiO$_3$\,[s]), forsterite (Mg$_2$SiO$_4$\,[s]), fayalite (Fe$_2$SiO$_4$\,[s]), olivine ((MgFe)$_2$SiO$_4$\,[s]), and quartz (SiO$_2$\,[s]). We also considered corundum (Al$_2$O$_3$\,[s]) as it was found to be the most likely dust species in the similar object KIC 1255b by \citet{Lieshout_2016}. The references for the dust optical data used are presented in Table~\ref{tab:dust}. Additionally, we used the reflectance/emissivity data from a large natural mineralogy library--JPL ECOSTRESS spectral library \citep{spectrallib}. 

Figure \ref{fig:solids} compares the spectrum of the fourth transit with the opacities for some of the main solid dust species we considered. The spectrum is inconsistent with featureless iron dust from a bare core (see Appendix  \ref{sec:app_4spectra}). 
The features may be consistent with some combination of solid mantle/crust mineralogy, but the significant transparency and opacity features seen at 4.4--6\,$\mu$m cannot be explained by any combination of the solid dust species we considered.  
        
        \begin{figure}[htb!]
        \centering
            \includegraphics[width=0.92\textwidth]{DACE_combined.png}
            \caption{Gas opacities of rock and ice vapor overlaid with the MIRI data. The two most theoretically plausible species with significant mid-infrared features that are expected from typical mantle material are shown: MgO and SiO \citep{Booth:2023:}. These poorly fit the data, suggesting that these gases are not detected in the MIRI data. Many different ice species, binned to the same resolution as the spectrum data, were considered from the DACE database. The best fitting gas models are nitric oxide, NO, and carbon dioxide, CO$_2$ to explain the short-wavelength features from 4.4--5.3\,$\mu$m.
            \label{fig:Gases}}
        \end{figure} 

        


\subsection{Gas Features} \label{sec:gas}   

The spectrum of the fourth transit, seen in Figure \ref{fig:solids}, shows a strikingly narrow feature around 5\,$\mu$m, where all solid models we considered exhibit smoothly continuous moderate opacity. This suggests the presence of a gas species. We searched the DACE database \citep{Grimm_2021} for comparable gas features and tested sets of species at high temperature (2100 K) and low pressure ($10^{-8}$ bar) that are expected for a vaporizing Earth-like rocky planet \citep[e.g.][]{Schaefer2009, Miguel2011}: CaO, TiO, AlO, MgO, SiO. We also examined related atomic species predicted by \citet{Booth:2023:}: Fe, Mg, O. Figure \ref{fig:Gases} shows the resulting opacities for the two most plausible of all of these gases, MgO and SiO, with significant features in mid-infrared. The poor fit to the data suggests these gases are not detected (as a dominant component). The optical opacity from the CHEOPS photometry measurements, despite large uncertainty, disfavors gas species with extreme optical opacity, such as Fe, Mg, O and MgO.

Since none of the expected rock vapor species were found to be good matches to the data, we also considered species known to be present in icy bodies in the solar system: CO, CO$_2$, NO, N$_2$, H$_2$O. The resulting opacities for the two most convincing of these species, NO and CO$_2$, can also be seen in Figure \ref{fig:Gases}. 
We note that the other ice species, including H$_2$O, are not ruled out at present. Contributions from those species are difficult to interpret with the data in hand, and a detailed analysis with more precise measurements is required to determine appropriate abundance ratios.
Finally, we considered more exotic rock-vapor species: FeH, SiS, SiN. None of the resulting opacities for these less plausible species matched the data.


\vspace{-0.5cm}

\section{Discussion} \label{sec:disc}


\subsection{Composition}

Assessing the habitability of rocky exoplanets requires knowledge of their interior compositions and mineralogy, which otherwise cannot be directly observed (although it can be inferred for some planets in the Solar System). 
Although USPs are undoubtedly inhospitable, they provide a unique case study for determining the compositions for a whole family of similar terrestrial planets and ascertaining deviations (or lack thereof) from their host-star compositions. 
Observations of USPs like K2-22b provide the opportunity to compare to observed stellar elemental abundances to constrain the interior compositions of more temperate extant rocky exoplanets.

Terrestrial rocky planets generally differentiate into metal-rich cores, rocky mantles, and chemically distinct crusts. Earth's mantle is silicate-rich (mostly MgSiO$_3$ with some small Fe/Mg ratio) \citep[e.g.][]{Kargel1993, BSE}. Thus, the most likely minerals to condense from a cloud of mantle vapor would be expected to be enstatite (MgSiO$_3$) and forsterite (Mg$_2$SiO$_4$). However, recent modeling by \citet{CamposEstrada:2024:} suggests a composition of magnesium–iron olivines or pyroxenes best explains the average transit shape of K2-22b. Similarly, \citet{Bromley:2023:} argued that iron-bearing silicates are required to be able to produce an observable mass loss rate for catastrophically evaporating exoplanets. Figure \ref{spectrum} demonstrates that improved long-wavelength (${\scriptstyle \gtrsim}8\mu$m) spectral fidelity is critical to testing these hypotheses. However, the features seen at these longer wavelengths are at least consistent with rocky silicate, as opposed to an iron core composition. 

Additionally, \citet{Booth:2023:} predicted opacity from likely gas species (SiO, MgO, Fe, O, and Mg) outgassing during the vaporization process, see Figure \ref{fig:Gases}. The shorter MIRI wavelength features in the middle panel spectrum in Figure \ref{spectrum} may constitute the first direct observations of gas features from an evaporating planet. Unexpectedly, the models that best fit these measurements seem to be typically ice-derived species (NO and CO$_2$, see Figure \ref{fig:Gases}). This is counter-intuitive to what we might expect from the evaporation of a rocky mantle and demands follow-up JWST observations for verification. 

One potential scenario to form these species could be photo-dissociation (X-ray and EUV) and collisional dissociation (via energetic electrons) of molecular nitrogen and water vapor (from water ice) \citep{marsh2004empirical, sinnhuber2012energetic, bulak2022quantification}. Alternatively, the NO can be produced by sublimation of H$_2$O and NH$_3$ ices followed by gas phase chemistry through OH and NH, a mechanism that has been partially used to explain NO observations in some proto-stellar objects \citep{Kulterer:2024:}. A potential geophysical situation wherein we might get the right combination of gas and dust species (as inferred from our JWST data analysis) is the evaporation of a mantle layer with some mixture of N$_2$-NH$_3$-CO$_2$ ice clathrates (at the base of a deep icy ocean world) \citep{Journaux:2020:, Carnahan:2022:} and silicate minerals. An alternative possibility is that a Ceres-like body is vaporizing, with a sufficient reservoir of carbon and nitrogen in organics and ammonium-rich ices/minerals \citep{diab2023bulk,nathues2024consus,de2024ammonium}. This seems even more plausible given the low mass of this object inferred from previous studies \citep{Schlawin:2021:}, ${\scriptstyle \lesssim}$0.03M$_\oplus$. 
Another possibility is magmatic degassing of N$_2$, CO$_2$, and H$_2$O dissolved in an outgassing silicate magma ocean \citep{chazallon2018selectivity, scarpa1996chemical,Kite:2016:}. The Earth's mantle is only 120 ppm C and 1--2 ppm N \citep{McDonough:1995:,johnson2015nitrogen}, suggesting that it may be difficult to get enough CO$_2$ and NO to explain our observations of K2-22b, assuming it has a bulk composition similar to Earth. A potentially mineralogical source for NO could be the decomposition of sinoite (${\rm Si}_{2}{\rm N}_{2}{\rm O}$), a mineral sometimes found in enstatite chondrites \citep{Keil_1965Natur.207..745K}, or nierite (${\rm Si}_{3}{\rm N}_{4}$), a mineral found in enstatite and ordinary chondrites \citep{Lee_1995Metic..30..387L}. However, these are rare mineral types overall and would require significant differences in mineralogical composition (e.g., high abundance of metal nitride) from what we might expect to see on Earth (e.g., in terms of planetary accretion and mantle redox state) \citep{mysen2019nitrogen}. Several of these scenarios imply vaporization of significant amounts of H$_2$O, the presence and abundance of which remains indeterminate without a more extensive analysis and more precise future measurements.

Confirming the spectral features seen for K2-22b with additional data from JWST is critical to better assess amongst the various plausible geophysical scenarios discussed above by quantitatively comparing observed gas abundances with thermo-chemical models with gas-phase photo-chemistry \citep{Curry2:2024:}. 

\subsection{Variability}

K2-22b's variable behavior has persisted well beyond the \textit{K2} mission \citep{colon2018large,Schlawin:2021:}, though recent observations indicate a long-term trend of quiescence \citep{Gaidos:2024:}. However, this is not entirely unprecedented, as the disintegrating planet KIC 1255b has been observed with a similar reduction in activity \citep{Schlawin:2016:}, only to return to its original activity levels a few years later \citep{Schlawin_KIC1255:2018:}. Several plausible interpretations for the long-term variability have been suggested: variations in stellar activity \citep{Kawahara:2013:}, changing interactions between the stellar radiation field and charged particles \citep{Mendis:2013:}, or substellar magma pool compositional variations \citep{Kite:2016:}. Long-term monitoring is required to provide observational constraints on the underlying mechanisms for this time evolution \citep{Gaidos:2024:}. 

Simulations of the chaotic outflows from this class of planet have been used to show that the dynamic evolution of the transiting dust cloud model is plausible over orbital timescales \citep{Bromley:2023:}. Hydrodynamical models of dust outflow geometry suggest K2-22b is a less massive planet than the similar object KIC 1255b, and particles escape the surface in a manner between the extremes of purely day-side and spherical outflow \citep{CamposEstrada:2024:}. There is likely a complex interplay between the optical depth of the resulting cloud and the particles that can further condense out and add to it \citep{Bromley:2023:}. Clouds of significantly differing optical depth may exhibit variations in abundance measurements of certain compositional elements. Inhomogeneities in the material being sublimated over time could impact the dominance of iron abundance in the cloud, interacting with the stellar magnetic field, and affecting cloud formation and dissipation rates.

Optical photometry is useful to directly sample the activity level of this object, and tease out the different timescales involved. However, additional spectral measurements with JWST are crucial for linking the activity cycles with compositional differences. Reassessing the statistical distribution of K2-22b transits at the current reduced activity levels seen in these results and \citet{Gaidos:2024:}, $\sim$28 additional observations of individual transits with out-of-transit baseline ($\sim$100 hours total) would offer $>$95\% confidence in obtaining the original 0.5--1\% cumulative transit depth sought after with our original JWST program. While it may be more economical to re-observe K2-22b with JWST only if and when it resumes higher activity levels, getting robust measurements during this quiescent phase could serve as an important point of comparison since the similarities and differences in the measured compositions between quiescent and active phases could inform us on the physics and chemistry behind the dominant source of variability. 

\subsection{Looking Forward}\label{sec:future}

The results of these observations have some interesting implications for the disintegrating planet model of K2-22b. The variability and spectral features remain consistent with the ephemeral nature of a transiting cloud of dust forming dynamically from the evaporating surface material of a closely orbiting body. 
However, if the short-wavelength features are indeed opacity from ``ice"-derived vapor species (in concert with mantle silicate components), then a terrestrial planet progenitor that formed within the snow line would be challenging to explain with traditional planet formation theories. Regardless of its composition, the close-in orbit seems plausibly explained by some migration pathway (e.g., stellar tides \citep{wu2024tidal}). This object must be re-observed to confirm the presence of these gas features, precisely measure the gas abundances, and determine an appropriate planetary migration and progenitor model. Additionally, getting detailed spectra during multiple phases of activity rates will constrain the impact of compositional heterogeneity on the observable variability. If these features can be confirmed in mid-IR, then there are likely near-IR spectral features as well that may warrant additional observations with NIRSpec. Given the overlap in wavelength range between NIRSpec and MIRI around 5\,$\mu$m, this may be an effective independent strategy to verify and identify additional gas features since molecules like water and CO$_2$ have strong NIR features \citep{Rothman:2009:}. NIRSpec observations may be particularly important for precise measurements given the additional uncertainty in the short-wavelength range of MIRI due to possible contamination from spectral foldover, as noted in Appendix \ref{sec:app_4spectra}.

The success of this experiment, despite the low precision of the measurements due to unexpectedly low activity rates, strongly validates this technique to explore the chemical composition of exoplanetary interiors and demonstrates the promise of using JWST to further explore K2-22b and other similar targets, such as KIC 1255b \citep{Rappaport_2012} and the newly discovered BD+05 4868 Ab \citep{TESS_Hon:2025:}\footnote{The fourth target in this category, KOI-2700b, is significantly fainter and likely not ideal for follow-up with JWST}. We have categorized all of these systems together as disintegrating exoplanets, but there are several observables that make K2-22b unique. 
K2-22 is the smallest/coolest host star, all of the others being mid-K-type stars, leading to a low $\beta$ parameter \citep{sanchis2015k2} (possibly of similar value to BD+05 4868 Ab, \citealt{TESS_Hon:2025:}) that influences the dynamics and evolution of the outflowing material. At $\sim$9 hours, K2-22b has the shortest period, allowing more observational opportunities than the other targets in a given time frame. The transit of K2-22b is more symmetric and shorter in duration than others, showing significant forward scattering in the \emph{K2} lightcurve without the gradually decreasing post-transit depression from an extended trailing tail. The newly discovered BD+05 4868 A is significantly brighter than the other stars hosting these planets and will soon be targeted for observations, but K2-22b offers a unique point of comparison considering the similarities and differences between the two objects.
Perhaps it is an outlier or perhaps we will find similar spectral features in all of these targets. By re-observing K2-22b and applying this technique to the other disintegrating planets, we can begin to understand how unique each system is, discover commonalities between them, and ultimately try to derive planetary interior parameters that we can trace back to the host star, which we can then apply to other systems.

\bigskip
\section*{ACKNOWLEDGEMENTS}

This work is based on observations made with the NASA/ESA/CSA James Webb Space Telescope. The data were obtained from MAST at STScI, which is operated by the Association of Universities for Research in Astronomy, Inc., under NASA contract NAS 5-03127 for JWST. These observations are associated with GO Program 3315.

Support for GO Program 3315 was provided by NASA through a grant from the Space Telescope Science Institute, which is operated by the Association of Universities for Research in Astronomy, Inc., under NASA contract NAS 5-03127.

This material is based upon work supported by the National Science Foundation Graduate Research Fellowship Program under Grant No. DGE1255832. 

The work of HPO has been carried out within the framework of the NCCR PlanetS supported by the Swiss National Science Foundation under grants 51NF40\_182901 and 51NF40\_205606.

The Center for Exoplanets and Habitable Worlds and the Penn State Extraterrestrial Intelligence Center are supported by Penn State and its Eberly College of Science.

CHEOPS is an ESA mission in partnership with Switzerland with important contributions to the payload and the ground segment from Austria, Belgium, France, Germany, Hungary, Italy, Portugal, Spain, Sweden, and the United Kingdom. The CHEOPS Consortium would like to gratefully acknowledge the support received by all the agencies, offices, universities, and industries involved. Their flexibility and willingness to explore new approaches were essential to the success of this mission. CHEOPS data analysed in this article will be made available in the CHEOPS mission archive (\href{https://cheops.unige.ch/archive_browser/}{https://cheops.unige.ch/archive\_browser/}).


\facilities{JWST (MIRI LRS), CHEOPS}

\software{Astropy \citep{astropy:2013, astropy:2018, astropy:2022},
          \texttt{Eureka!}\ \citep{Eureka_Bell2022}
          , \texttt{Pegasus}}

\clearpage
\bibliography{K2-22b_references}{}

\begin{thebibliography}{}
\expandafter\ifx\csname natexlab\endcsname\relax\def\natexlab#1{#1}\fi
\providecommand{\url}[1]{\href{#1}{#1}}
\providecommand{\dodoi}[1]{doi:~\href{http://doi.org/#1}{\nolinkurl{#1}}}
\providecommand{\doeprint}[1]{\href{http://ascl.net/#1}{\nolinkurl{http://ascl.net/#1}}}
\providecommand{\doarXiv}[1]{\href{https://arxiv.org/abs/#1}{\nolinkurl{https://arxiv.org/abs/#1}}}

\bibitem[{{Adibekyan} {et~al.}(2021){Adibekyan}, {Dorn}, {Sousa}, {Santos}, {Bitsch}, {Israelian}, {Mordasini}, {Barros}, {Delgado Mena}, {Demangeon}, {Faria}, {Figueira}, {Hakobyan}, {Oshagh}, {Soares}, {Kunitomo}, {Takeda}, {Jofr{\'e}}, {Petrucci}, \& {Martioli}}]{Adibekyan:2021:}
{Adibekyan}, V., {Dorn}, C., {Sousa}, S.~G., {et~al.} 2021, Science, 374, 330, \dodoi{10.1126/science.abg8794}

\bibitem[{{Astropy Collaboration} {et~al.}(2013){Astropy Collaboration}, {Robitaille}, {Tollerud}, {Greenfield}, {Droettboom}, {Bray}, {Aldcroft}, {Davis}, {Ginsburg}, {Price-Whelan}, {Kerzendorf}, {Conley}, {Crighton}, {Barbary}, {Muna}, {Ferguson}, {Grollier}, {Parikh}, {Nair}, {Unther}, {Deil}, {Woillez}, {Conseil}, {Kramer}, {Turner}, {Singer}, {Fox}, {Weaver}, {Zabalza}, {Edwards}, {Azalee Bostroem}, {Burke}, {Casey}, {Crawford}, {Dencheva}, {Ely}, {Jenness}, {Labrie}, {Lim}, {Pierfederici}, {Pontzen}, {Ptak}, {Refsdal}, {Servillat}, \& {Streicher}}]{astropy:2013}
{Astropy Collaboration}, {Robitaille}, T.~P., {Tollerud}, E.~J., {et~al.} 2013, \aap, 558, A33, \dodoi{10.1051/0004-6361/201322068}

\bibitem[{{Astropy Collaboration} {et~al.}(2018){Astropy Collaboration}, {Price-Whelan}, {Sip{\H{o}}cz}, {G{\"u}nther}, {Lim}, {Crawford}, {Conseil}, {Shupe}, {Craig}, {Dencheva}, {Ginsburg}, {Vand erPlas}, {Bradley}, {P{\'e}rez-Su{\'a}rez}, {de Val-Borro}, {Aldcroft}, {Cruz}, {Robitaille}, {Tollerud}, {Ardelean}, {Babej}, {Bach}, {Bachetti}, {Bakanov}, {Bamford}, {Barentsen}, {Barmby}, {Baumbach}, {Berry}, {Biscani}, {Boquien}, {Bostroem}, {Bouma}, {Brammer}, {Bray}, {Breytenbach}, {Buddelmeijer}, {Burke}, {Calderone}, {Cano Rodr{\'\i}guez}, {Cara}, {Cardoso}, {Cheedella}, {Copin}, {Corrales}, {Crichton}, {D'Avella}, {Deil}, {Depagne}, {Dietrich}, {Donath}, {Droettboom}, {Earl}, {Erben}, {Fabbro}, {Ferreira}, {Finethy}, {Fox}, {Garrison}, {Gibbons}, {Goldstein}, {Gommers}, {Greco}, {Greenfield}, {Groener}, {Grollier}, {Hagen}, {Hirst}, {Homeier}, {Horton}, {Hosseinzadeh}, {Hu}, {Hunkeler}, {Ivezi{\'c}}, {Jain}, {Jenness}, {Kanarek}, {Kendrew}, {Kern}, {Kerzendorf}, {Khvalko}, {King}, {Kirkby}, {Kulkarni},
  {Kumar}, {Lee}, {Lenz}, {Littlefair}, {Ma}, {Macleod}, {Mastropietro}, {McCully}, {Montagnac}, {Morris}, {Mueller}, {Mumford}, {Muna}, {Murphy}, {Nelson}, {Nguyen}, {Ninan}, {N{\"o}the}, {Ogaz}, {Oh}, {Parejko}, {Parley}, {Pascual}, {Patil}, {Patil}, {Plunkett}, {Prochaska}, {Rastogi}, {Reddy Janga}, {Sabater}, {Sakurikar}, {Seifert}, {Sherbert}, {Sherwood-Taylor}, {Shih}, {Sick}, {Silbiger}, {Singanamalla}, {Singer}, {Sladen}, {Sooley}, {Sornarajah}, {Streicher}, {Teuben}, {Thomas}, {Tremblay}, {Turner}, {Terr{\'o}n}, {van Kerkwijk}, {de la Vega}, {Watkins}, {Weaver}, {Whitmore}, {Woillez}, {Zabalza}, \& {Astropy Contributors}}]{astropy:2018}
{Astropy Collaboration}, {Price-Whelan}, A.~M., {Sip{\H{o}}cz}, B.~M., {et~al.} 2018, \aj, 156, 123, \dodoi{10.3847/1538-3881/aabc4f}

\bibitem[{{Astropy Collaboration} {et~al.}(2022){Astropy Collaboration}, {Price-Whelan}, {Lim}, {Earl}, {Starkman}, {Bradley}, {Shupe}, {Patil}, {Corrales}, {Brasseur}, {N{"o}the}, {Donath}, {Tollerud}, {Morris}, {Ginsburg}, {Vaher}, {Weaver}, {Tocknell}, {Jamieson}, {van Kerkwijk}, {Robitaille}, {Merry}, {Bachetti}, {G{"u}nther}, {Aldcroft}, {Alvarado-Montes}, {Archibald}, {B{'o}di}, {Bapat}, {Barentsen}, {Baz{'a}n}, {Biswas}, {Boquien}, {Burke}, {Cara}, {Cara}, {Conroy}, {Conseil}, {Craig}, {Cross}, {Cruz}, {D'Eugenio}, {Dencheva}, {Devillepoix}, {Dietrich}, {Eigenbrot}, {Erben}, {Ferreira}, {Foreman-Mackey}, {Fox}, {Freij}, {Garg}, {Geda}, {Glattly}, {Gondhalekar}, {Gordon}, {Grant}, {Greenfield}, {Groener}, {Guest}, {Gurovich}, {Handberg}, {Hart}, {Hatfield-Dodds}, {Homeier}, {Hosseinzadeh}, {Jenness}, {Jones}, {Joseph}, {Kalmbach}, {Karamehmetoglu}, {Ka{l}uszy{'n}ski}, {Kelley}, {Kern}, {Kerzendorf}, {Koch}, {Kulumani}, {Lee}, {Ly}, {Ma}, {MacBride}, {Maljaars}, {Muna}, {Murphy}, {Norman}, {O'Steen},
  {Oman}, {Pacifici}, {Pascual}, {Pascual-Granado}, {Patil}, {Perren}, {Pickering}, {Rastogi}, {Roulston}, {Ryan}, {Rykoff}, {Sabater}, {Sakurikar}, {Salgado}, {Sanghi}, {Saunders}, {Savchenko}, {Schwardt}, {Seifert-Eckert}, {Shih}, {Jain}, {Shukla}, {Sick}, {Simpson}, {Singanamalla}, {Singer}, {Singhal}, {Sinha}, {Sip{H{o}}cz}, {Spitler}, {Stansby}, {Streicher}, {{{S}}umak}, {Swinbank}, {Taranu}, {Tewary}, {Tremblay}, {Val-Borro}, {Van Kooten}, {Vasovi{'c}}, {Verma}, {de Miranda Cardoso}, {Williams}, {Wilson}, {Winkel}, {Wood-Vasey}, {Xue}, {Yoachim}, {Zhang}, {Zonca}, \& {Astropy Project Contributors}}]{astropy:2022}
{Astropy Collaboration}, {Price-Whelan}, A.~M., {Lim}, P.~L., {et~al.} 2022, \apj, 935, 167, \dodoi{10.3847/1538-4357/ac7c74}

\bibitem[{{Ballmer} \& {Noack}(2021)}]{Ballmer:2021:}
{Ballmer}, M.~D., \& {Noack}, L. 2021, Elements, 17, 245, \dodoi{10.2138/gselements.17.4.245}

\bibitem[{{Beatty} {et~al.}(2024){Beatty}, {Welbanks}, {Schlawin}, {Bell}, {Line}, {Murphy}, {Edelman}, {Greene}, {Fortney}, {Henry}, {Mukherjee}, {Ohno}, {Parmentier}, {Rauscher}, {Wiser}, \& {Arnold}}]{Beatty2024}
{Beatty}, T.~G., {Welbanks}, L., {Schlawin}, E., {et~al.} 2024, \apjl, 970, L10, \dodoi{10.3847/2041-8213/ad55e9}

\bibitem[{{Begemann} {et~al.}(1997){Begemann}, {Dorschner}, {Henning}, {Mutschke}, {G{\"u}rtler}, {K{\"o}mpe}, \& {Nass}}]{AlO3_Begemann}
{Begemann}, B., {Dorschner}, J., {Henning}, T., {et~al.} 1997, \apj, 476, 199, \dodoi{10.1086/303597}

\bibitem[{Bell {et~al.}(2022)Bell, Ahrer, Brande, Carter, Feinstein, {Guzman Caloca}, Mansfield, Zieba, Piaulet, Benneke, Filippazzo, May, Roy, Kreidberg, \& Stevenson}]{Eureka_Bell2022}
Bell, T.~J., Ahrer, E.-M., Brande, J., {et~al.} 2022, Journal of Open Source Software, 7, 4503, \dodoi{10.21105/joss.04503}

\bibitem[{{Bell} {et~al.}(2023){Bell}, {Kreidberg}, {Kendrew}, {Bean}, {Crouzet}, {Ducrot}, {Dyrek}, {Gao}, {Lagage}, \& {Moses}}]{Bell_2023arXiv230106350B}
{Bell}, T.~J., {Kreidberg}, L., {Kendrew}, S., {et~al.} 2023, arXiv e-prints, arXiv:2301.06350, \dodoi{10.48550/arXiv.2301.06350}

\bibitem[{Benz {et~al.}(2021)Benz, Broeg, Fortier, Rando, Beck, Beck, Queloz, Ehrenreich, Maxted, Isaak, {et~al.}}]{benz2021cheops}
Benz, W., Broeg, C., Fortier, A., {et~al.} 2021, Experimental Astronomy, 51, 109

\bibitem[{Bodman {et~al.}(2018)Bodman, Wright, Desch, \& Lisse}]{bodman2018inferring}
Bodman, E.~H., Wright, J.~T., Desch, S.~J., \& Lisse, C.~M. 2018, The Astronomical Journal, 156, 173

\bibitem[{{Booth} {et~al.}(2023){Booth}, {Owen}, \& {Schulik}}]{Booth:2023:}
{Booth}, R.~A., {Owen}, J.~E., \& {Schulik}, M. 2023, \mnras, 518, 1761, \dodoi{10.1093/mnras/stac3121}

\bibitem[{{Bouchet} {et~al.}(2022){Bouchet}, {Gastaud}, {Lagage}, {Kendrew}, {Bombardi}, {Coulais}, {Ronayette}, {Sloan}, {Moreau}, {Orduna}, {Gr{\'e}goire}, {Dyrek}, {Bouwman}, {Glasse}, \& {Wright}}]{Bouchet:2022:}
{Bouchet}, P., {Gastaud}, R., {Lagage}, P.~O., {et~al.} 2022, in Society of Photo-Optical Instrumentation Engineers (SPIE) Conference Series, Vol. 12180, Space Telescopes and Instrumentation 2022: Optical, Infrared, and Millimeter Wave, ed. L.~E. {Coyle}, S.~{Matsuura}, \& M.~D. {Perrin}, 121800Z, \dodoi{10.1117/12.2629778}

\bibitem[{{Bouwman} {et~al.}(2023){Bouwman}, {Kendrew}, {Greene}, {Bell}, {Lagage}, {Schreiber}, {Dicken}, {Sloan}, {Espinoza}, {Scheithauer}, {Coulais}, {Fox}, {Gastaud}, {Glauser}, {Jones}, {Labiano}, {Lahuis}, {Morrison}, {Murray}, {Mueller}, {Nayak}, {Wright}, {Glasse}, \& {Rieke}}]{Bouwman:2023:}
{Bouwman}, J., {Kendrew}, S., {Greene}, T.~P., {et~al.} 2023, \pasp, 135, 038002, \dodoi{10.1088/1538-3873/acbc49}

\bibitem[{{Bromley} \& {Chiang}(2023)}]{Bromley:2023:}
{Bromley}, J., \& {Chiang}, E. 2023, \mnras, 521, 5746, \dodoi{10.1093/mnras/stad932}

\bibitem[{Bulak {et~al.}(2022)Bulak, Paardekooper, Fedoseev, Chuang, van Scheltinga, Eistrup, \& Linnartz}]{bulak2022quantification}
Bulak, M., Paardekooper, D., Fedoseev, G., {et~al.} 2022, Astronomy \& Astrophysics, 657, A120

\bibitem[{{Campos Estrada} {et~al.}(2024){Campos Estrada}, {Owen}, {Jankovic}, {Wilson}, \& {Helling}}]{CamposEstrada:2024:}
{Campos Estrada}, B., {Owen}, J.~E., {Jankovic}, M.~R., {Wilson}, A., \& {Helling}, C. 2024, \mnras, 528, 1249, \dodoi{10.1093/mnras/stae095}

\bibitem[{{Carnahan} {et~al.}(2022){Carnahan}, {Vance}, {Hesse}, {Journaux}, \& {Sotin}}]{Carnahan:2022:}
{Carnahan}, E., {Vance}, S.~D., {Hesse}, M.~A., {Journaux}, B., \& {Sotin}, C. 2022, \grl, 49, e97602, \dodoi{10.1029/2021GL097602}

\bibitem[{Chazallon \& Pirim(2018)}]{chazallon2018selectivity}
Chazallon, B., \& Pirim, C. 2018, Chemical Engineering Journal, 342, 171

\bibitem[{{Cockell} {et~al.}(2024){Cockell}, {Simons}, {Castillo-Rogez}, {Higgins}, {Kaltenegger}, {Keane}, {Leonard}, {Mitchell}, {Park}, {Perl}, \& {Vance}}]{Cockell:2024:}
{Cockell}, C.~S., {Simons}, M., {Castillo-Rogez}, J., {et~al.} 2024, Nature Astronomy, 8, 675, \dodoi{10.1038/s41550-024-02196-w}

\bibitem[{Col{\'o}n {et~al.}(2018)Col{\'o}n, Zhou, Shporer, Collins, Bieryla, Espinoza, Murgas, Pattarakijwanich, Awiphan, Armstrong, {et~al.}}]{colon2018large}
Col{\'o}n, K.~D., Zhou, G., Shporer, A., {et~al.} 2018, The Astronomical Journal, 156, 227

\bibitem[{{Curry} {et~al.}(2024{\natexlab{a}}){Curry}, {Booth}, {Owen}, \& {Mohanty}}]{Curry1_interior_model:2024:}
{Curry}, A., {Booth}, R., {Owen}, J.~E., \& {Mohanty}, S. 2024{\natexlab{a}}, \mnras, 528, 4314, \dodoi{10.1093/mnras/stae191}

\bibitem[{{Curry} {et~al.}(2024{\natexlab{b}}){Curry}, {Booth}, {Owen}, \& {Mohanty}}]{Curry2:2024:}
---. 2024{\natexlab{b}}, \mnras, 528, 4314, \dodoi{10.1093/mnras/stae191}

\bibitem[{De~Sanctis {et~al.}(2024)De~Sanctis, Ammannito, Carrozzo, Ciarniello, De~Angelis, Ferrari, Frigeri, \& Raponi}]{de2024ammonium}
De~Sanctis, M.~C., Ammannito, E., Carrozzo, F., {et~al.} 2024, Communications Earth \& Environment, 5, 131

\bibitem[{Diab {et~al.}(2023)Diab, Daswani, \& Castillo-Rogez}]{diab2023bulk}
Diab, J., Daswani, M.~M., \& Castillo-Rogez, J. 2023, Icarus, 391, 115339

\bibitem[{{Dorn} {et~al.}(2015){Dorn}, {Khan}, {Heng}, {Connolly}, {Alibert}, {Benz}, \& {Tackley}}]{Dorn:2015:}
{Dorn}, C., {Khan}, A., {Heng}, K., {et~al.} 2015, \aap, 577, A83, \dodoi{10.1051/0004-6361/201424915}

\bibitem[{Egger {et~al.}(2024)Egger, Osborn, Kubyshkina, Mordasini, Alibert, G{\"u}nther, Lendl, Brandeker, Heitzmann, Leleu, {et~al.}}]{egger2024unveiling}
Egger, J., Osborn, H., Kubyshkina, D., {et~al.} 2024, Astronomy \& Astrophysics, 688, A223

\bibitem[{{Foley}(2024)}]{Foley:2024:}
{Foley}, B.~J. 2024, Reviews in Mineralogy and Geochemistry, 90, 559, \dodoi{10.2138/rmg.2024.90.15}

\bibitem[{Fortier {et~al.}(2024)Fortier, Simon, Broeg, Olofsson, Deline, Wilson, Maxted, Brandeker, Cameron, Beck, {et~al.}}]{fortier2024cheops}
Fortier, A., Simon, A., Broeg, C., {et~al.} 2024, arXiv preprint arXiv:2406.01716

\bibitem[{{Gaidos} {et~al.}(2024){Gaidos}, {Parviainen}, {Esparza-Borges}, {Fukui}, {Isogai}, {Kawauchi}, {de Leon}, {Mori}, {Murgas}, {Narita}, {Palle}, \& {Watanabe}}]{Gaidos:2024:}
{Gaidos}, E., {Parviainen}, H., {Esparza-Borges}, E., {et~al.} 2024, \aap, 688, L34, \dodoi{10.1051/0004-6361/202451332}

\bibitem[{{Grimm} {et~al.}(2021){Grimm}, {Malik}, {Kitzmann}, {Guzm{\'a}n-Mesa}, {Hoeijmakers}, {Fisher}, {Mendon{\c{c}}a}, {Yurchenko}, {Tennyson}, {Alesina}, {Buchschacher}, {Burnier}, {Segransan}, {Kurucz}, \& {Heng}}]{Grimm_2021}
{Grimm}, S.~L., {Malik}, M., {Kitzmann}, D., {et~al.} 2021, \apjs, 253, 30, \dodoi{10.3847/1538-4365/abd773}

\bibitem[{{Henning} {et~al.}(1995){Henning}, {Begemann}, {Mutschke}, \& {Dorschner}}]{FeO_Henning}
{Henning}, T., {Begemann}, B., {Mutschke}, H., \& {Dorschner}, J. 1995, \aaps, 112, 143

\bibitem[{{Henning} \& {Mutschke}(1997)}]{FeS_Henning}
{Henning}, T., \& {Mutschke}, H. 1997, \aap, 327, 743

\bibitem[{{Hon} {et~al.}(2025){Hon}, {Rappaport}, {Shporer}, {Vanderburg}, {Collins}, {Watkins}, {Schwarz}, {Barkaoui}, {Yee}, {Winn}, {Polanski}, {Gilbert}, {Ciardi}, {Audenaert}, {Fong}, {Haviland}, {Hesse}, {Muthukrishna}, {Petitpas}, {Hadjiyska Schmelzer}, {Narita}, {Fukui}, {Seager}, \& {Ricker}}]{TESS_Hon:2025:}
{Hon}, M., {Rappaport}, S., {Shporer}, A., {et~al.} 2025, arXiv e-prints, arXiv:2501.05431, \dodoi{10.48550/arXiv.2501.05431}

\bibitem[{{Hughes} {et~al.}(2018){Hughes}, {Duch{\^e}ne}, \& {Matthews}}]{Hughes:2018:}
{Hughes}, A.~M., {Duch{\^e}ne}, G., \& {Matthews}, B.~C. 2018, \araa, 56, 541, \dodoi{10.1146/annurev-astro-081817-052035}

\bibitem[{{J{\"a}ger} {et~al.}(2003){J{\"a}ger}, {Dorschner}, {Mutschke}, {Posch}, \& {Henning}}]{Mg2SiO4_Jager}
{J{\"a}ger}, C., {Dorschner}, J., {Mutschke}, H., {Posch}, T., \& {Henning}, T. 2003, \aap, 408, 193, \dodoi{10.1051/0004-6361:20030916}

\bibitem[{Johnson \& Goldblatt(2015)}]{johnson2015nitrogen}
Johnson, B., \& Goldblatt, C. 2015, Earth-Science Reviews, 148, 150

\bibitem[{{Journaux} {et~al.}(2020){Journaux}, {Kalousov{\'a}}, {Sotin}, {Tobie}, {Vance}, {Saur}, {Bollengier}, {Noack}, {R{\"u}ckriemen-Bez}, {Van Hoolst}, {Soderlund}, \& {Brown}}]{Journaux:2020:}
{Journaux}, B., {Kalousov{\'a}}, K., {Sotin}, C., {et~al.} 2020, \ssr, 216, 7, \dodoi{10.1007/s11214-019-0633-7}

\bibitem[{{Jura} \& {Young}(2014)}]{Jura2014}
{Jura}, M., \& {Young}, E.~D. 2014, Annual Review of Earth and Planetary Sciences, 42, 45, \dodoi{10.1146/annurev-earth-060313-054740}

\bibitem[{{Kargel} \& {Lewis}(1993)}]{Kargel1993}
{Kargel}, J.~S., \& {Lewis}, J.~S. 1993, \icarus, 105, 1, \dodoi{10.1006/icar.1993.1108}

\bibitem[{{Kawahara} {et~al.}(2013){Kawahara}, {Hirano}, {Kurosaki}, {Ito}, \& {Ikoma}}]{Kawahara:2013:}
{Kawahara}, H., {Hirano}, T., {Kurosaki}, K., {Ito}, Y., \& {Ikoma}, M. 2013, \apjl, 776, L6, \dodoi{10.1088/2041-8205/776/1/L6}

\bibitem[{{Keil}(1965)}]{Keil_1965Natur.207..745K}
{Keil}, K. 1965, \nat, 207, 745, \dodoi{10.1038/207745a0}

\bibitem[{{Kite} {et~al.}(2016){Kite}, {Fegley}, {Schaefer}, \& {Gaidos}}]{Kite:2016:}
{Kite}, E.~S., {Fegley}, Bruce, J., {Schaefer}, L., \& {Gaidos}, E. 2016, \apj, 828, 80, \dodoi{10.3847/0004-637X/828/2/80}

\bibitem[{{Kitzmann} \& {Heng}(2018)}]{Kitzmann2018}
{Kitzmann}, D., \& {Heng}, K. 2018, \mnras, 475, 94, \dodoi{10.1093/mnras/stx3141}

\bibitem[{{Koike} {et~al.}(1995){Koike}, {Kaito}, {Yamamoto}, {Shibai}, {Kimura}, \& {Suto}}]{Al2O3_Koike}
{Koike}, C., {Kaito}, C., {Yamamoto}, T., {et~al.} 1995, \icarus, 114, 203, \dodoi{10.1006/icar.1995.1055}

\bibitem[{{Kulterer} {et~al.}(2024){Kulterer}, {Wampfler}, {Ligterink}, {Murillo}, {Hsieh}, {McClure}, {Boogert}, {Kipfer}, {Bjerkeli}, \& {Drozdovskaya}}]{Kulterer:2024:}
{Kulterer}, B.~M., {Wampfler}, S.~F., {Ligterink}, N.~F.~W., {et~al.} 2024, \aap, 691, A281, \dodoi{10.1051/0004-6361/202450792}

\bibitem[{{Laor} \& {Draine}(1993)}]{Laor1993}
{Laor}, A., \& {Draine}, B.~T. 1993, \apj, 402, 441, \dodoi{10.1086/172149}

\bibitem[{{Lee} {et~al.}(1995){Lee}, {Russell}, {Arden}, \& {Pillinger}}]{Lee_1995Metic..30..387L}
{Lee}, M.~R., {Russell}, S.~S., {Arden}, J.~W., \& {Pillinger}, C.~T. 1995, Meteoritics, 30, 387, \dodoi{10.1111/j.1945-5100.1995.tb01142.x}

\bibitem[{Marsh {et~al.}(2004)Marsh, Solomon, \& Reynolds}]{marsh2004empirical}
Marsh, D., Solomon, S., \& Reynolds, A. 2004, Journal of Geophysical Research: Space Physics, 109

\bibitem[{{Matsuo} {et~al.}(2019){Matsuo}, {Greene}, {Johnson}, {Mcmurray}, {Roellig}, \& {Ennico}}]{Matsuo_2019}
{Matsuo}, T., {Greene}, T.~P., {Johnson}, R.~R., {et~al.} 2019, \pasp, 131, 124502, \dodoi{10.1088/1538-3873/ab42f1}

\bibitem[{{McDonough} \& {Sun}(1995)}]{McDonough:1995:}
{McDonough}, W.~F., \& {Sun}, S.~s. 1995, Chemical Geology, 120, 223, \dodoi{10.1016/0009-2541(94)00140-4}

\bibitem[{{Meerdink} {et~al.}(2019){Meerdink}, {Hook}, {Roberts}, \& {Abbott}}]{spectrallib}
{Meerdink}, S.~K., {Hook}, S.~J., {Roberts}, D.~A., \& {Abbott}, E.~A. 2019, Remote Sensing of Environment, 230, 111196, \dodoi{10.1016/j.rse.2019.05.015}

\bibitem[{{Mendis} \& {Hor{\'a}nyi}(2013)}]{Mendis:2013:}
{Mendis}, D.~A., \& {Hor{\'a}nyi}, M. 2013, Reviews of Geophysics, 51, 53, \dodoi{10.1002/rog.20005}

\bibitem[{{Mie}(1908)}]{Mie1908}
{Mie}, G. 1908, Annalen der Physik, 330, 377, \dodoi{10.1002/andp.19083300302}

\bibitem[{{Miguel} {et~al.}(2011){Miguel}, {Kaltenegger}, {Fegley}, \& {Schaefer}}]{Miguel2011}
{Miguel}, Y., {Kaltenegger}, L., {Fegley}, B., \& {Schaefer}, L. 2011, \apjl, 742, L19, \dodoi{10.1088/2041-8205/742/2/L19}

\bibitem[{Mysen(2019)}]{mysen2019nitrogen}
Mysen, B. 2019, Progress in Earth and Planetary Science, 6, 1

\bibitem[{Nathues {et~al.}(2024)Nathues, Hoffmann, Sarkar, Singh, Hernandez, Pasckert, Schmedemann, Thangjam, Cloutis, Mengel, {et~al.}}]{nathues2024consus}
Nathues, A., Hoffmann, M., Sarkar, R., {et~al.} 2024, Journal of Geophysical Research: Planets, 129, e2023JE008150

\bibitem[{O'Neill \& Palme(1998)}]{BSE}
O'Neill, H. S.~C., \& Palme, H. 1998, Composition of the Silicate Earth: Implications for Accretion and Core Formation (Cambridge University Press), 3 -- 126, \dodoi{10.1017/CBO9780511573101.004}

\bibitem[{{Palik}(1985)}]{Palik1985}
{Palik}, E.~D. 1985, {Handbook of optical constants of solids}

\bibitem[{{Palik}(1991)}]{Palik1991}
---. 1991, {Handbook of optical constants of solids II}

\bibitem[{{Perez-Becker}(2013)}]{Perez-Becker_2013PhDT.......300P}
{Perez-Becker}, D.~A. 2013, PhD thesis, University of California, Berkeley

\bibitem[{{Pollack} {et~al.}(1994){Pollack}, {Hollenbach}, {Beckwith}, {Simonelli}, {Roush}, \& {Fong}}]{FeS_Pollack}
{Pollack}, J.~B., {Hollenbach}, D., {Beckwith}, S., {et~al.} 1994, \apj, 421, 615, \dodoi{10.1086/173677}

\bibitem[{{Posch} {et~al.}(2003){Posch}, {Kerschbaum}, {Fabian}, {Mutschke}, {Dorschner}, {Tamanai}, \& {Henning}}]{CaTiO3_Posch}
{Posch}, T., {Kerschbaum}, F., {Fabian}, D., {et~al.} 2003, \apjs, 149, 437, \dodoi{10.1086/379167}

\bibitem[{{Rappaport} {et~al.}(2014){Rappaport}, {Barclay}, {DeVore}, {Rowe}, {Sanchis-Ojeda}, \& {Still}}]{Rappaport:2014:}
{Rappaport}, S., {Barclay}, T., {DeVore}, J., {et~al.} 2014, \apj, 784, 40, \dodoi{10.1088/0004-637X/784/1/40}

\bibitem[{Rappaport {et~al.}(2012)Rappaport, Levine, Chiang, El~Mellah, Jenkins, Kalomeni, Kite, Kotson, Nelson, Rousseau-Nepton, \& et~al.}]{Rappaport_2012}
Rappaport, S., Levine, A., Chiang, E., {et~al.} 2012, The Astrophysical Journal, 752, 1, \dodoi{10.1088/0004-637x/752/1/1}

\bibitem[{{Rothman} {et~al.}(2009){Rothman}, {Gordon}, {Barbe}, {Benner}, {Bernath}, {Birk}, {Boudon}, {Brown}, {Campargue}, {Champion}, {Chance}, {Coudert}, {Dana}, {Devi}, {Fally}, {Flaud}, {Gamache}, {Goldman}, {Jacquemart}, {Kleiner}, {Lacome}, {Lafferty}, {Mandin}, {Massie}, {Mikhailenko}, {Miller}, {Moazzen-Ahmadi}, {Naumenko}, {Nikitin}, {Orphal}, {Perevalov}, {Perrin}, {Predoi-Cross}, {Rinsland}, {Rotger}, {{\v{S}}ime{\v{c}}kov{\'a}}, {Smith}, {Sung}, {Tashkun}, {Tennyson}, {Toth}, {Vandaele}, \& {Vander Auwera}}]{Rothman:2009:}
{Rothman}, L.~S., {Gordon}, I.~E., {Barbe}, A., {et~al.} 2009, \jqsrt, 110, 533, \dodoi{10.1016/j.jqsrt.2009.02.013}

\bibitem[{Sanchis-Ojeda {et~al.}(2015)Sanchis-Ojeda, Rappaport, Pall{\`e}, Delrez, DeVore, Gandolfi, Fukui, Ribas, Stassun, Albrecht, {et~al.}}]{sanchis2015k2}
Sanchis-Ojeda, R., Rappaport, S., Pall{\`e}, E., {et~al.} 2015, The Astrophysical Journal, 812, 112

\bibitem[{Scarpa {et~al.}(1996)Scarpa, Tilling, \& Giggenbach}]{scarpa1996chemical}
Scarpa, R., Tilling, R.~I., \& Giggenbach, W. 1996, Monitoring and mitigation of volcano hazards, 221

\bibitem[{{Schaefer} \& {Fegley}(2009)}]{Schaefer2009}
{Schaefer}, L., \& {Fegley}, B. 2009, \apjl, 703, L113, \dodoi{10.1088/0004-637X/703/2/L113}

\bibitem[{{Schlawin} {et~al.}(2016){Schlawin}, {Herter}, {Zhao}, {Teske}, \& {Chen}}]{Schlawin:2016:}
{Schlawin}, E., {Herter}, T., {Zhao}, M., {Teske}, J.~K., \& {Chen}, H. 2016, \apj, 826, 156, \dodoi{10.3847/0004-637X/826/2/156}

\bibitem[{{Schlawin} {et~al.}(2018){Schlawin}, {Hirano}, {Kawahara}, {Teske}, {Green}, {Rackham}, {Fraine}, \& {Bushra}}]{Schlawin_KIC1255:2018:}
{Schlawin}, E., {Hirano}, T., {Kawahara}, H., {et~al.} 2018, \aj, 156, 281, \dodoi{10.3847/1538-3881/aaeb32}

\bibitem[{{Schlawin} {et~al.}(2021){Schlawin}, {Su}, {Herter}, {Ridden-Harper}, \& {Apai}}]{Schlawin:2021:}
{Schlawin}, E., {Su}, K. Y.~L., {Herter}, T., {Ridden-Harper}, A., \& {Apai}, D. 2021, \aj, 162, 57, \dodoi{10.3847/1538-3881/ac0b41}

\bibitem[{{Schlawin} {et~al.}(2024{\natexlab{a}}){Schlawin}, {Mukherjee}, {Ohno}, {Bell}, {Beatty}, {Greene}, {Line}, {Challener}, {Parmentier}, {Fortney}, {Rauscher}, {Wiser}, {Welbanks}, {Murphy}, {Edelman}, {Batalha}, {Moran}, {Mehta}, \& {Rieke}}]{Schalwin2024a}
{Schlawin}, E., {Mukherjee}, S., {Ohno}, K., {et~al.} 2024{\natexlab{a}}, \aj, 168, 104, \dodoi{10.3847/1538-3881/ad58e0}

\bibitem[{{Schlawin} {et~al.}(2024{\natexlab{b}}){Schlawin}, {Ohno}, {Bell}, {Murphy}, {Welbanks}, {Beatty}, {Greene}, {Fortney}, {Parmentier}, {Edelman}, {Gill}, {Anderson}, {Wheatley}, {Henry}, {Mehta}, {Kreidberg}, \& {Rieke}}]{Schalwin2024b}
{Schlawin}, E., {Ohno}, K., {Bell}, T.~J., {et~al.} 2024{\natexlab{b}}, \apjl, 974, L33, \dodoi{10.3847/2041-8213/ad7fef}

\bibitem[{{Seager} {et~al.}(2007){Seager}, {Kuchner}, {Hier-Majumder}, \& {Militzer}}]{Seager:2007:}
{Seager}, S., {Kuchner}, M., {Hier-Majumder}, C.~A., \& {Militzer}, B. 2007, \apj, 669, 1279, \dodoi{10.1086/521346}

\bibitem[{Sinnhuber {et~al.}(2012)Sinnhuber, Nieder, \& Wieters}]{sinnhuber2012energetic}
Sinnhuber, M., Nieder, H., \& Wieters, N. 2012, Surveys in Geophysics, 33, 1281

\bibitem[{{Swain} {et~al.}(2024){Swain}, {Hasegawa}, {Thorngren}, \& {Roudier}}]{Swain:2024:}
{Swain}, M.~R., {Hasegawa}, Y., {Thorngren}, D.~P., \& {Roudier}, G.~M. 2024, \ssr, 220, 61, \dodoi{10.1007/s11214-024-01098-7}

\bibitem[{{Thiabaud} {et~al.}(2015){Thiabaud}, {Marboeuf}, {Alibert}, {Leya}, \& {Mezger}}]{Thiabaud:2015:}
{Thiabaud}, A., {Marboeuf}, U., {Alibert}, Y., {Leya}, I., \& {Mezger}, K. 2015, \aap, 580, A30, \dodoi{10.1051/0004-6361/201525963}

\bibitem[{{Ueda} {et~al.}(1998){Ueda}, {Yanagi}, {Noshiro}, {Hosono}, \& {Kawazoe}}]{CaTiO3_Ueda}
{Ueda}, K., {Yanagi}, H., {Noshiro}, R., {Hosono}, H., \& {Kawazoe}, H. 1998, Journal of Physics Condensed Matter, 10, 3669, \dodoi{10.1088/0953-8984/10/16/018}

\bibitem[{{van Lieshout} {et~al.}(2014){van Lieshout}, {Min}, \& {Dominik}}]{Lieshout_2014}
{van Lieshout}, R., {Min}, M., \& {Dominik}, C. 2014, \aap, 572, A76, \dodoi{10.1051/0004-6361/201424876}

\bibitem[{{van Lieshout} {et~al.}(2016){van Lieshout}, {Min}, {Dominik}, {Brogi}, {de Graaff}, {Hekker}, {Kama}, {Keller}, {Ridden-Harper}, \& {van Werkhoven}}]{Lieshout_2016}
{van Lieshout}, R., {Min}, M., {Dominik}, C., {et~al.} 2016, \aap, 596, A32, \dodoi{10.1051/0004-6361/201629250}

\bibitem[{{Vanderburg} {et~al.}(2015){Vanderburg}, {Johnson}, {Rappaport}, {Bieryla}, {Irwin}, {Lewis}, {Kipping}, {Brown}, {Dufour}, {Ciardi}, {Angus}, {Schaefer}, {Latham}, {Charbonneau}, {Beichman}, {Eastman}, {McCrady}, {Wittenmyer}, \& {Wright}}]{Vanderburg2015}
{Vanderburg}, A., {Johnson}, J.~A., {Rappaport}, S., {et~al.} 2015, \nat, 526, 546, \dodoi{10.1038/nature15527}

\bibitem[{{Wang} {et~al.}(2019){Wang}, {Liu}, {Ireland}, {Brasser}, {Yong}, \& {Lineweaver}}]{Wang:2019:}
{Wang}, H.~S., {Liu}, F., {Ireland}, T.~R., {et~al.} 2019, \mnras, 482, 2222, \dodoi{10.1093/mnras/sty2749}

\bibitem[{{Welbanks} {et~al.}(2024){Welbanks}, {Bell}, {Beatty}, {Line}, {Ohno}, {Fortney}, {Schlawin}, {Greene}, {Rauscher}, {McGill}, {Murphy}, {Parmentier}, {Tang}, {Edelman}, {Mukherjee}, {Wiser}, {Lagage}, {Dyrek}, \& {Arnold}}]{Welbanks2024}
{Welbanks}, L., {Bell}, T.~J., {Beatty}, T.~G., {et~al.} 2024, \nat, 630, 836, \dodoi{10.1038/s41586-024-07514-w}

\bibitem[{Wu {et~al.}(2024)Wu, Dewberry, \& Fuller}]{wu2024tidal}
Wu, S.~C., Dewberry, J.~W., \& Fuller, J. 2024, The Astrophysical Journal, 963, 34

\bibitem[{{Zeidler} {et~al.}(2011){Zeidler}, {Posch}, {Mutschke}, {Richter}, \& {Wehrhan}}]{Zeidler2011}
{Zeidler}, S., {Posch}, T., {Mutschke}, H., {Richter}, H., \& {Wehrhan}, O. 2011, \aap, 526, A68, \dodoi{10.1051/0004-6361/201015219}

\end{thebibliography}
\bibliographystyle{aasjournal}

\appendix
\restartappendixnumbering
\vspace{-0.5cm}
\section{Data Reduction}\label{sec:reduction}

    \subsection{Eureka!} \label{sec:eureka}

        We reduced the Stage 0 uncal JWST data products from MAST using the \texttt{Eureka!}\footnote{https://github.com/kevin218/Eureka!}\ pipeline with the standard configuration files for the MIRI instrument \citep{Eureka_Bell2022}. We tested a range of apertures to account for background contamination from the companion star. Using the apertures that produced the minimum noise profiles, we ran Stage 1 to Stage 4 to obtain time series flux measurements over each observed exposure in 300 spectral channels from 4.4 to 11.8\,$\mu$m. The same data reduction process was used for all 4 observations, with the only difference being the optimal apertures selected for each observation. We used vertical axis rows 150 to 393 along the dispersion axis, and horizontal axis columns 10 to 67. For transits 1 \& 2, we selected 15 pixels for the half-width aperture around K2-22 and 25 pixels as the start of the background aperture to effectively mask out the companion. For transits 3 \& 4, we used half-width apertures of 6 and 19 for the target and background, respectively.
        
            \begin{figure}[htb!]
                \centering
                \includegraphics[width=0.92\textwidth]{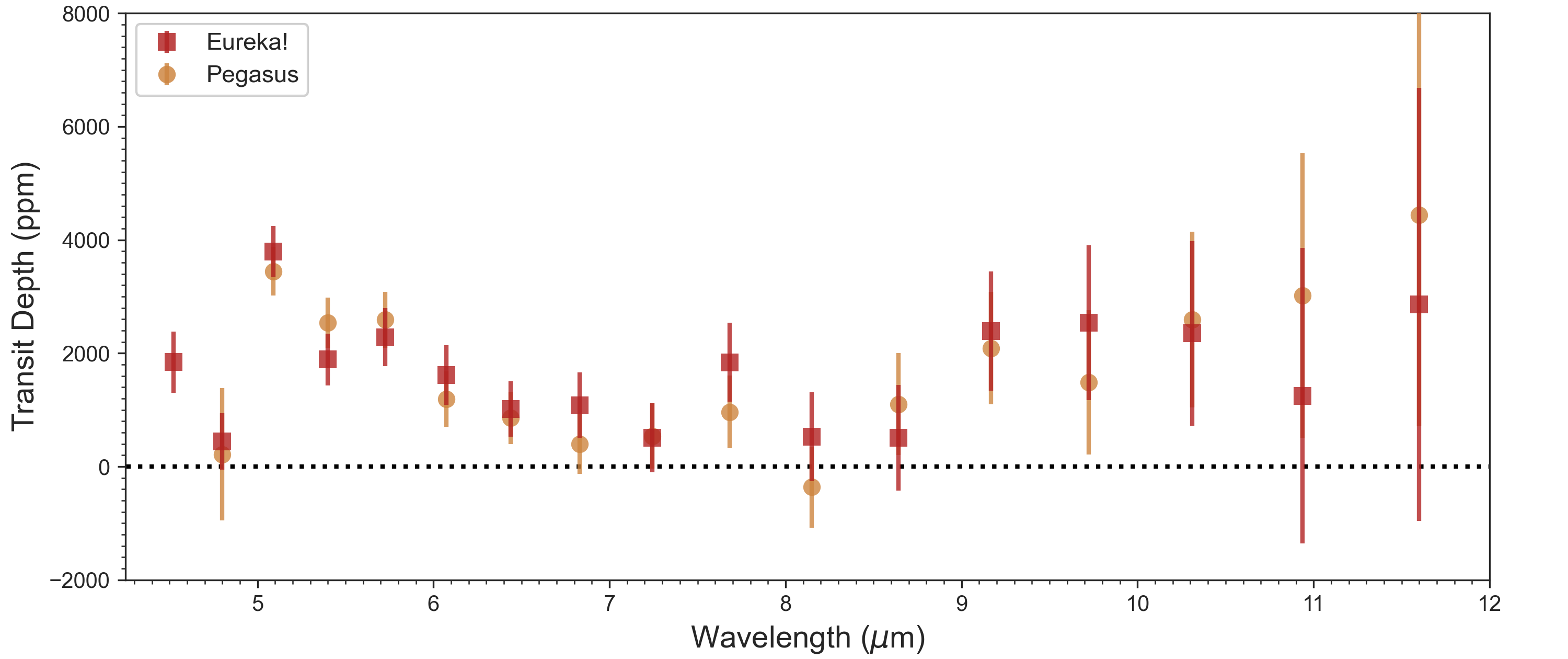}
                \caption{Two independent reductions of the MIRI LRS data for transit 4 agree within their uncertainties. Shown are our fiducial \texttt{Eureka!}\ results (Section \ref{sec:eureka}) and the results from a reduction using the \texttt{Pegasus} pipeline (Section \ref{sec:pegasus}), plotted with their associated $1\,\sigma$ uncertainties at a constant spectral resolution of $R=17$. Both reductions agree on the overall shape of the transmission spectrum and are consistent with each other to within $1\,\sigma$.}
                \label{fig:reductions}
            \end{figure}
        
\medskip            
    \subsection{Pegasus} \label{sec:pegasus}

        As an independent check on the \texttt{Eureka!}\ data reduction described in Section \ref{sec:eureka}, we also reduced the observations using the \texttt{Pegasus} pipeline\footnote{\url{https://github.com/TGBeatty/PegasusProject}}, which has been used for NIRCam \citep{Beatty2024,Welbanks2024,Schalwin2024a} and NIRSpec \citep{Schalwin2024b} transit and eclipse observations. Our reduction began with background subtraction on the \textsc{rateints} files provided by version 1.13.3 of the \texttt{jwst} pipeline using CRDS version 11.17.14. We first performed a general background subtraction step for each \textsc{rateints} file by fitting a two-dimensional, second-order spline to each integration using the entire 72$\times$416 \textsc{rateints} images. For the background fitting, we masked columns 0 to 12 and 67 to 723 to remove the non-imaged sections of the subarray. To prevent self-subtraction, we also masked out starlight from K2-22 by masking pixels in columns 29 to 45 and above row 50, and from the companion star by masking the pixels in columns 15 to 25 and above row 150. We then performed a single round of $3\,\sigma$ clipping on the unmasked portions of the image. We then fit a spline to the unmasked, unclipped pixel values with a median-box size of 5 pixels. We extrapolated the combined background spline for the whole image over the masked portions near K2-22 and the companion star and subtracted it from the original image values. We then removed the alternating row and column horizontal and vertical banding in the background-subtracted images by calculating the biweight mean of each row or column, and then subtracting this mean from each row or column. In both cases, we used the same starlight-mask as for the spline fitting to avoid self-subtraction.

        We then extracted broadband and spectroscopic lightcurves from our background-subtracted images. To do so, we fit the spectral trace using a fourth-order polynomial and then used optimal extraction to measure the 1D spectrum in each image. We performed three rounds of iterative profile estimation for the optimal extraction routine, after which we judged the profile fit to have converged. Using the resultant 1D spectra, we extracted a broadband lightcurve from 4.65\,$\mu$m to 11.95\,$\mu$m. For the spectroscopic lightcurves, we subdivided this wavelength region into 17 spectral channels spaced at a constant spectral resolution of $R=17$. For both the broadband and spectroscopic lightcurves, we linearly interpolated over each spectral column to account for partial-pixel effects in the wavelength solutions.

        Figure \ref{fig:reductions} shows that independent reduction pipelines produce good agreement of the observed transit spectrum.
\medskip
    \subsection{CHEOPS Data Reduction} \label{sec:CHEOPS_reduction}

        To extract precise photometry and remove systematics (e.g. due to field rotation) we used PSF photometric extraction using the \texttt{PIPE} package \citep[see e.g.][]{fortier2024cheops}\footnote{https://github.com/alphapsa/PIPE}. We then performed a combined model using \texttt{cheoxplanet} \citep[see e.g.][]{egger2024unveiling}\footnote{https://github.com/hposborn/chexoplanet}. 
        In order to remove coherent variations caused by spacecraft rotation, an array of 24 independent flux parameters spaced evenly in roll angle (average spacing 9 degrees) is included in the model and a cubic spline drawn from these points is subtracted from the lightcurve.

\bigskip
\section{Outlier Removal} \label{sec:app_outliers}

        \begin{figure}[htb!]
            \centering
            \includegraphics[width=0.92\textwidth]{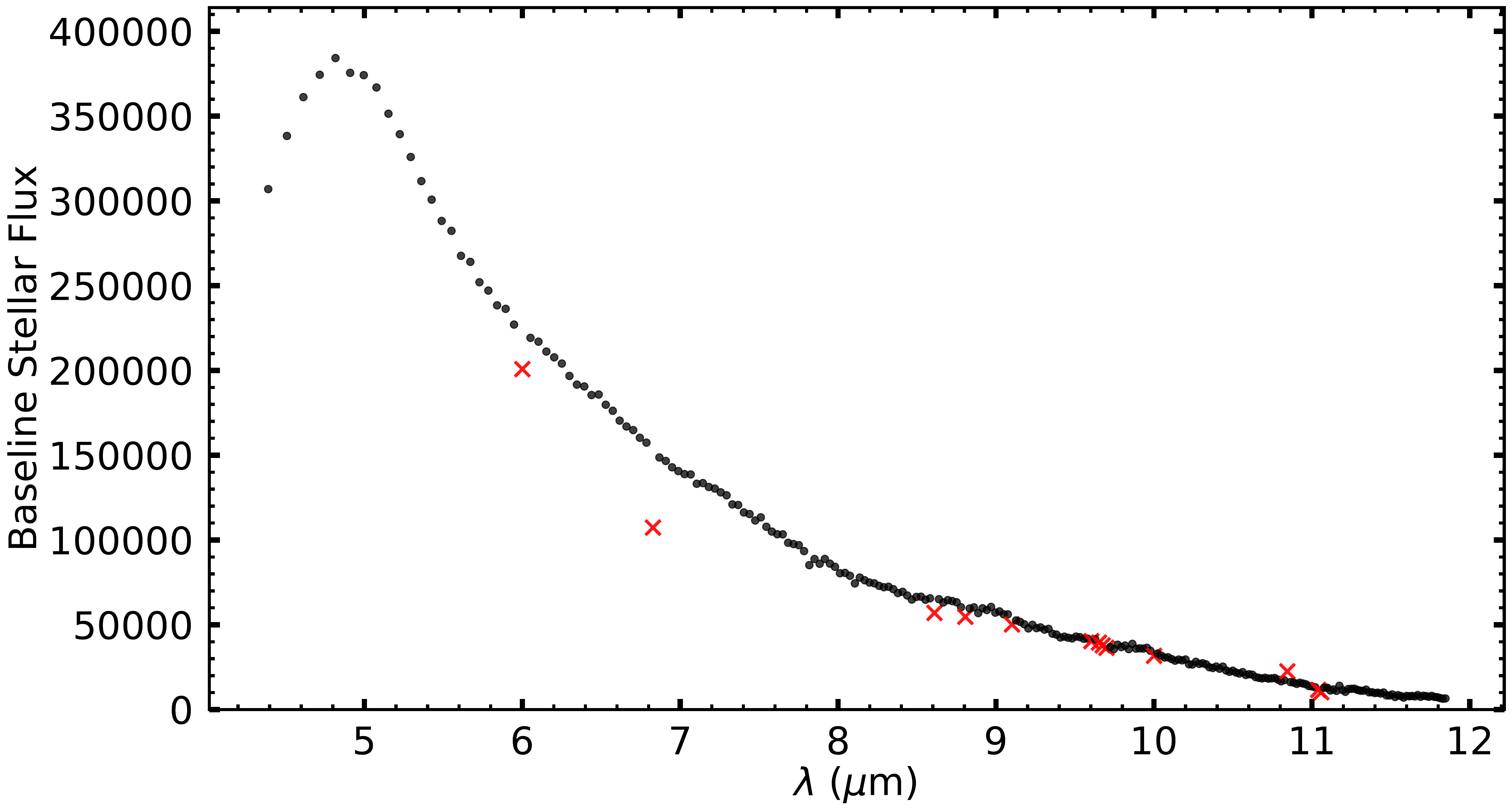}
            \caption{Time-averaged stellar spectrum with flux outliers removed. Turnover below 5\,$\mu$m is due to reduced throughput. A sliding window of 9 channels was used to identify points with a greater than 3$\sigma$ deviation from its neighbors. Red ``x" points are the outliers that were removed. 
            \label{fig:outliers}}
        \end{figure}

Figure \ref{fig:outliers} shows the stellar spectrum from MIRI, time averaged over the fourth observation. The turnover below 5\,$\mu$m is due to reduced throughput in slitless mode. The red ``x" points indicate outliers in the spectrum that deviated from neighboring points in a 9-channel wide sliding window by 3$\sigma$. These outliers were removed from the data as a first step before subsequent analysis.

\section{Lightcurve Analysis} 
\label{sec:app_lightcurves}

    The target of these observations is an ephemeral cloud of dust and gas, rapidly evolving over the timescale of a single orbit ($\sim$9 h), and the shape and depth of an individual transit deviates unpredictably from the classic solid-body transit model. 
    Due to the unique nature of this object, subsequent analysis of the data was done using bespoke python code to normalize and model the lightcurve transit. The time series output from \texttt{Eureka!}\ was curated by removing the first 30 minutes of data to account for instrument settling \citep{Bell_2023arXiv230106350B,Bouwman:2023:,Matsuo_2019}, and then removing $>3\sigma$ outliers from the stellar spectrum (see Figure \ref{fig:outliers}).

        \begin{figure}[htb!]
            \centering
            \includegraphics[width=0.92\textwidth]{lightcurves_shortof8.png}
            \caption{MIRI LRS lightcurves for all 4 transit observations listed in Tables \ref{tab:obs} \& \ref{tab:results} for combined light of the spectral channels up to 8\,$\mu$m. The unbinned data (at a cadence of 72 seconds)are greyed out in the background. The blue points are binned to a time resolution of 8 minutes. The red line is a 2nd order polynomial fit to transit 4 and scaled to the best fit value in each window. Multiple transit models were fit, including the original \textit{K2} lightcurve, a skewed Gaussian, and a boxcar, with no significant difference in the overall depth or resulting spectrum. The 46 min transit duration calculated by \citet{sanchis2015k2} is included as vertical gray dashed lines for reference. The uncertainties in the data points as reported by \texttt{Eureka!} were inflated to match the actual standard deviation in the baseline flux. 
            \label{fig:all4LCs}}
        \end{figure}

        \begin{figure}[htb!]
            \centering
            \includegraphics[width=0.92\textwidth]{Allan_combined_light_shortof8_T4.png}
            \caption{A standard assessment of the residuals of the model fit to the lightcurve for transit 4 using the combined light of the MIRI spectral channels up to 8\,$\mu$m. The blue line shows the RMS of the residuals of the transit model fit to the data for different bin sizes. The unbinned data was taken with a cadence of 72 seconds, providing roughly 38 points across the assumed average transit duration. The orange line shows how our adopted single measurement uncertainty would scale as white noise as the data are binned up. The uncertainties in the data points as reported by \texttt{Eureka!}\ were inflated by a factor to match the actual standard deviation in the baseline flux for more conservatively accurate error estimates while preserving variations in the measurements.
            \label{fig:allan}}
        \end{figure}

        \begin{figure}[htb!]
            \centering
            \includegraphics[width=0.92\textwidth]{LookElsewhere_short.png}
            \includegraphics[width=0.92\textwidth]{lookelsewhere_dist_split_short.png}
            \caption{\emph{Top:} Measured depth using the transit model as a fixed-duration sliding window across the combined-light time series for each observation up to 8\,$\mu$m. This shows that there is some variation in the normalized flux for each transit, possibly due to systematics from the MIRI instrument. The most significant measurement occurs at the predicted transit center time of the 4th transit window, while low significance measurements in transit windows 2 and 3 correspond to those reported in Section \ref{sec:results}. 
            \\\hspace{\textwidth}
            \emph{Bottom:} Histogram of depth measurements (blue) at all times outside 1.5$\times$ the 46min duration of each transit window, i.e. ``elsewhere" in the observations. The standalone observations, 1 and 2, show distinctly increased uncertainties and are thus treated separately from the full phase curve. The standard deviations in the distributions are ${\scriptstyle \lesssim}$1.1$\times$ the calculated depth uncertainty, suggesting minimal effects from red noise. The red dashed line shows the measured depth at the center time of the 4th transit window for comparison.}
            \label{fig:lookelsewhere}
        \end{figure}
        
        \begin{figure}[htb!]
            \centering
            \includegraphics[width=0.85\textwidth]{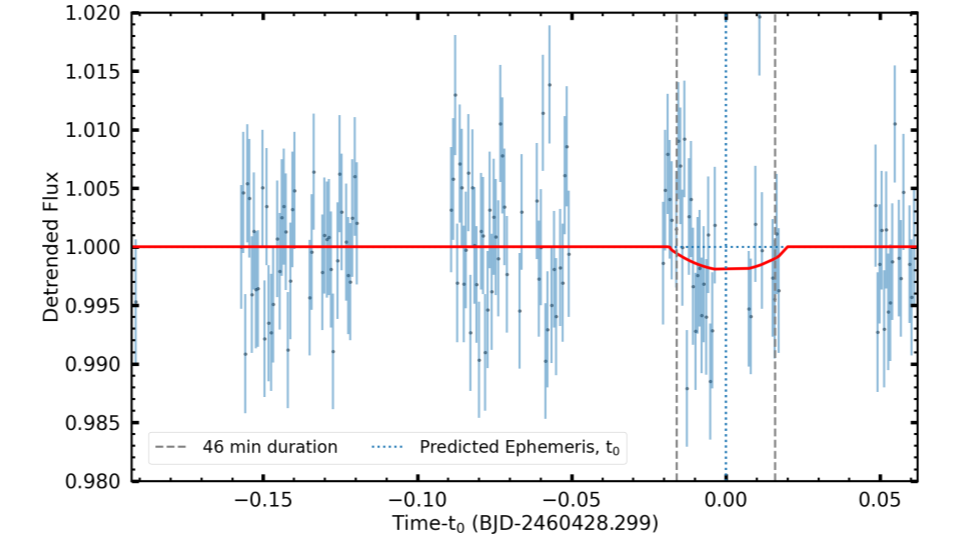}
            \caption{Detrended CHEOPS photometry during transit 4. The red line is a 2nd order polynomial fit to the transit window and zero elsewhere. Multiple transit models were fit, including the original \textit{K2} lightcurve, a skewed Gaussian, and a boxcar, with no significant difference. The 46min transit duration calculated by \citet{sanchis2015k2} is included for reference. Gaps in the data were caused by occultations of Earth during the space telescope's orbit. The other three transit windows fall in these gaps. 
            \label{fig:CHEOPS}}
        \end{figure}
    
    Using the cleaned data, we first plotted the white lightcurve for each observation and identified a clear transit during the fourth expected transit window. Focusing first on the most significant detection, we modeled the transit in white light, and then applied this model to each channel in order to measure the spectroscopic depth. The primary model chosen is a simple 2-degree polynomial over the transit window that goes to zero at the window edges. The average lightcurve from the original \textit{K2} data, a skewed Gaussian, and a simple boxcar were also tested as models, resulting in no significant difference in the derived spectrum, verifying that our results are not sensitive to this choice.
    
    The model depth, $\delta$, was calculated from

    \vspace{-0.2cm}
    \begin{equation}\label{eq:depth}
        \delta = \frac{\sum_i (y_i-1) f_i / \sigma_i^2}{\sum_i f_i^2 / \sigma_i^2},     
    \end{equation}
    where $y$ is the measured normalized flux, $f$ is the model, and $\sigma$ is the uncertainty in each data point, as measured by the standard deviation of the baseline flux. Due to the unknown extent of the transiting cloud, the baseline flux was conservatively chosen outside $1.5\times$ the 46 minute transit duration window calculated by \citet{sanchis2015k2} and centered at the ephemeris from \citet{Schlawin:2021:}. The model depth variance, $\sigma^2_\delta$, is calculated as
    
    \vspace{-0.2cm}
    \begin{equation}\label{eq:unc}
        \sigma_\delta^2 = \frac{1}{\sum_i f_i^2 / \sigma_i^2}.   
    \end{equation}

    We then applied these same techniques to the lightcurves and spectra of the other observations. 

    Figure \ref{fig:all4LCs} shows the cleaned time series data for all four transit observations described in Section \ref{sec:obs} and listed in Table \ref{tab:results}, for the combined light of the MIRI spectral channels up to 8\,$\mu$m. The first 30 minutes of data was removed from each observation to account for settling. The baseline of each observation was conservatively estimated as the flux outside $1.5\times$ the 46min average transit duration determined by \citet{sanchis2015k2}. For observations 1, 2 and 3, a line was fit to the baseline flux and divided out to detrend any residual ramp remaining after the \texttt{Eureka!}\ pipeline reduction. The reported uncertainties from \texttt{Eureka!}\ were inflated to match the standard deviation in the baseline flux. By increasing the reported uncertainties, as opposed to simply using the standard deviation in the data, we hoped to preserve any variations due to systematics that were properly accounted for in the reduction pipeline. We applied this same treatment of each time series in both the combined-light and the individual spectral channels for each observation. 
    
    Multiple bin sizes were tried, and we ultimately settled on binning to a time resolution of 8 minutes (N=21). Figure \ref{fig:allan} shows the results of plotting the standard deviation of the residuals of the model fit versus bin size, illustrating the absence of significant red noise. We constructed a lightcurve model by fitting a 2-degree polynomial to these binned up points in the transit window of transit 4, and setting the model to the mean of the baseline flux elsewhere. This model was then applied to each transit observation, centered around the ephemeris and scaled to measure the transit depth. These measurements are listed in Table \ref{tab:results}. The residuals for these models is shown in the bottom panels of each lightcurve in Figure \ref{fig:all4LCs}. Other lightcurve models were also tested, including the average \textit{K2} lightcurve, a skewed Gaussian, and a simple boxcar. The resulting spectra from fitting these models to the spectroscopic channels showed no significant difference, indicating our results are not sensitive to our choice of lightcurve model. 

    The ``look elsewhere" plots in Figure \ref{fig:lookelsewhere} show the results of taking the transit model, fixed in duration and allowing the depth to vary, and sliding it across the time series of the combined light of the MIRI spectral channels up to 8\,$\mu$m for each observation to measure the depth at every point. Equations \ref{eq:depth} \& \ref{eq:unc} are used to calculate the depth and uncertainty at each point. With tighter uncertainties and larger deviations from the normalized flux, the most significant depth measurement occurs during the 4th predicted transit window. These results also lend additional support for the low significance detections during the 2nd and 3rd transit windows. To account for differences between the isolated observations and the full phase curve, which are evident in the different noise profiles, we treated the distributions separately to examine the red noise. The standard deviation in the measured depths elsewhere is ${\scriptstyle \lesssim}$1.1$\times$ the uncertainty in our reported measurements listed in Table \ref{tab:results}, consistent with the absence of red noise shown in Figure \ref{fig:allan}. We therefore ignore the contribution of red noise in our analysis as a neglible effect. 
    These results support our report of a transit detection during the 4th transit window. 

    The same lightcurve models were tested on the CHEOPS photometry, also showing no significant difference in the results. However, 3 of the 4 transits of K2-22b were impacted by earth occultation gaps in our observing windows. Only the 4th transit window, and fortuitously the most significant transit event, lined up with an unobstructed observing window for CHEOPS. Figure \ref{fig:CHEOPS} shows the fit of our chosen 2-degree polynomial model to the detrended CHEOPS data over the transit window. We measured the depth and associated uncertainty of each transit fit using equations \ref{eq:depth} \& \ref{eq:unc}. The magnitude of K2-22b ($G = 14.93$) is technically beyond the rated ($G = 6 - 12$) magnitude limit of CHEOPS \citep{fortier2024cheops}. The result was a photometric lightcurve with an RMS of 2371 ppm per 30-min bin. Given the CHEOPS bandpass of 330--1100 nm, we added this photometric point to our JWST spectrum as an optical depth constraint.

\section{Supporting Evidence for Spectral Features} \label{sec:app_4spectra}

    The resulting spectrum appears consistent with some combination of solid mineralogy over the wavelength range of 4.4--8\,$\mu$m with additional gas opacity to explain the short wavelength features (4.4--5.3\,$\mu$m), see Sections \ref{sec:solids} \& \ref{sec:gas}. Due to the large uncertainties at the longer wavelengths, it is difficult to differentiate between the different solid mineralogies. However, Figure \ref{fig:BIC} shows the results of calculations of a Bayesian information criterion (BIC) for a select group of mineralogy using two parameters to scale and offset the models as compared with a flat line with just one parameter. There is a positive significance for some mantle-like silicate mineralogy over an iron-dominated core-like mineralogy or flat spectrum. We performed the same test over the entire 4.4--11.8\,$\mu$m range, which favors the magnesium-rich silicates, while still disfavoring the iron-dominated core-like species. From this we conclude that our spectrum is more consistent with mantle mineralogy features than a featureless flat spectrum.
        
        \begin{figure}[htb!]
            \centering
            \includegraphics[width=0.85\textwidth]{BIC_2p_short.png}
            \caption{Calculated $\Delta$BIC values for select solid mineralogy opacities noted in Section \ref{sec:solids} and shown in Figure \ref{fig:solids} against a flat line as compared with our resulting spectrum at wavelengths $<$8\,$\mu$m. The top four models shown here, quartz (SiO$_2$ [s]), olivine ((MgFe)$_2$SiO$_4$ [s]), enstatite (MgSiO$_3$ [s]), and forsterite (Mg$_2$SiO$_4$ [s]), represent likely crust/mantle minerals. The bottom three, iron (Fe [s]), iron sulfide (FeS [s]), and iron oxide (FeO [s]), represent likely core minerals. We used 2 parameters to scale and offset the mineralogy models, and 1 parameter for the flat line. These $\Delta$ BIC values show that the spectrum data is more consistent with mantle mineralogy than either a flat spectrum or iron-dominant mineralogy. The different colored bars are a qualitative comparison of the results; Negative values are shown in red, positive values below 3 are colored orange, and those above 3 are green.}
            \label{fig:BIC}
        \end{figure}

    The short wavelength channels are prone to some amount of contamination due to spectral foldover from the LRS disperser element, a double prism assembly \citep{Bouchet:2022:}. The lowest bin in our JWST R=17 spectrum at 4.52\,$\mu$m contains spectral channels 4.39, 4.51, and 4.61\,$\mu$m.
    Figure 10 from \citet{Bouchet:2022:} shows that below 4.5\,$\mu$m there is significant contamination from the flux in channels blue-ward of the foldover point at 3.9\,$\mu$m. We estimate that the lowest bin in our JWST spectrum, containing the three shortest wavelength channels, could be contaminated by up to $\sim$30\% from flux between 3.25--3.5\,$\mu$m. Our next shortest bin, indicating significant transparency and including channels 4.72, 4.82, and 4.91\,$\mu$m, could be contaminated by up to $\sim$11\% from flux between 3.0--3.25\,$\mu$m. The most interesting and significant feature at 5.1\,$\mu$m includes channels 5.00, 5.08, and 5.15\,$\mu$m, and could be contaminated by up to $\sim$6\% from wavelengths less than 3\,$\mu$m. 
    A more precise evaluation of the strength of the features that correspond to our data would necessitate a more rigorous analysis of the potential contamination from spectral foldover and subsequent increased uncertainties. 
    However, spectral foldover is unlikely to significantly affect our value of the opacity in the 4.52\,$\mu$m bin, since the signal at 3.25--3.5\,$\mu$m is being diluted by a factor of at least 3.
    This further motivates the necessity to validate these short wavelength features with follow-up measurements, particularly with overlapping wavelength coverage in near-IR.

        \begin{figure}[htb!]
            \centering            
            \includegraphics[width=0.92\textwidth]{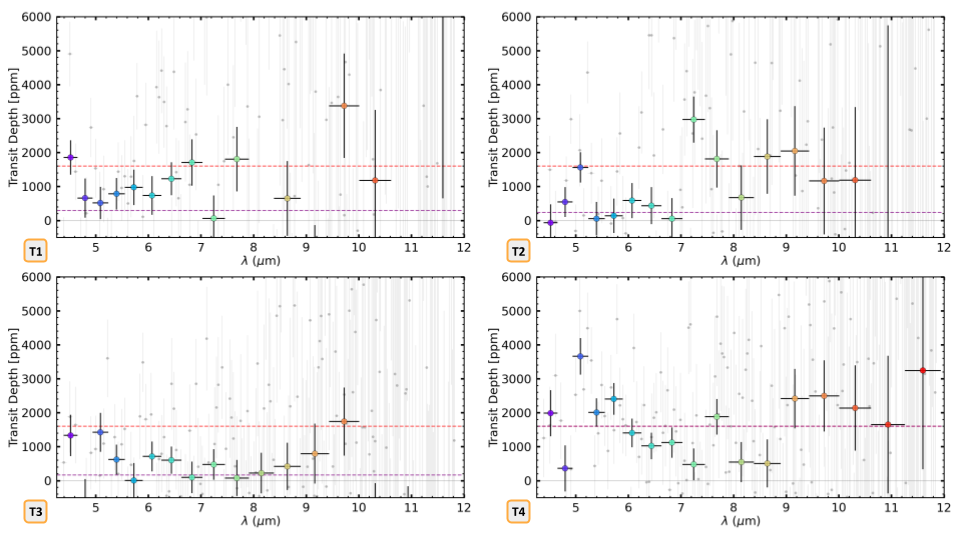}
            \centering
            \includegraphics[width=0.92\textwidth]{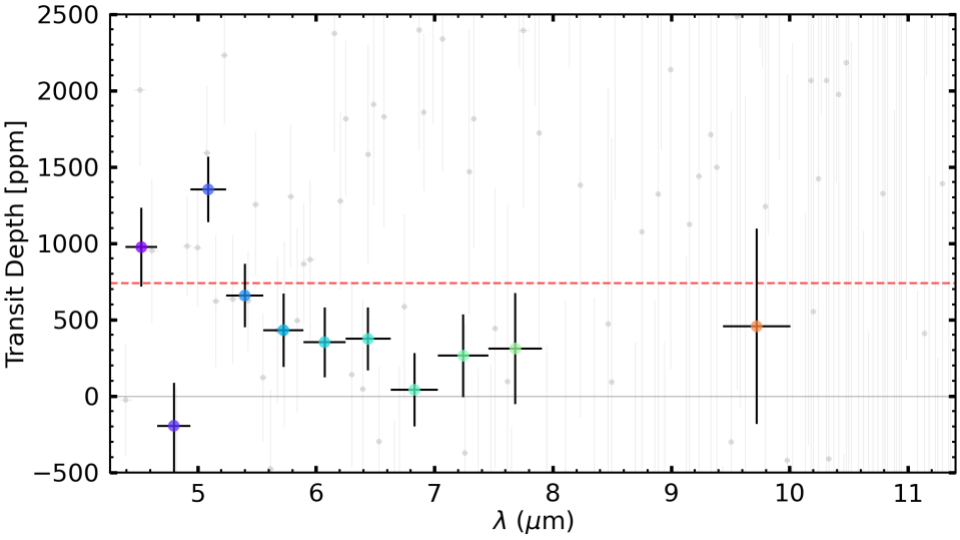}
            \caption{MIRI spectra during all four expected transits using the 2-degree transit model from transit 4, scaled to the optimal depth in each channel for each transit. 
            \\\hspace{\textwidth}
            \emph{Top:} Individual spectra during each transit window. The red dashed line marks the overall transit depth in T4. The purple dashed line indicates the overall transit depth in each observation as measured from the white-light lightcurves. The white-light time series of transit 4 shows a clear signal, while the others do not (see Figure \ref{fig:all4LCs}). 
            However, the most or second most significant detections in T1, T2 and T3 are in the 4.5 and 5.1\,$\mu$m channels, supporting the reality of these narrow features observed at modest significance in T4.
            \\\hspace{\textwidth}
            \emph{Bottom:} Combined spectra of all four transits. The 4.5 and 5.1\,$\mu$m features are clearly evident.
            \label{fig:4spectra}}
        \end{figure}

    Our assessment of the short-wavelength features is discussed in Section \ref{sec:gas}, and here we offer additional support for our results. The white-light time series for the observation \#1 shows no transit, and observations \#2 and \#3 show only low significance ($<$3-$\sigma$) transits in combined light over the reduced wavelength range of 4.4--8\,$\mu$m. Despite little to no evidence for white-light transit detections in these observations, we produced a spectrum for each transit window by fitting our transit model to each spectral channel. The top panel of Figure \ref{fig:4spectra} shows the resulting fit for each observation. The data is binned the same for each observation (R=17) and shows significant depth in the 4.5 and 5.1\,$\mu$m channels in the transits 1, 2, 3, despite their low overall white-light transit depth. And in every observation, the spectra show transparency in the 4.8\,$\mu$m bin.

    The bottom panel of Figure \ref{fig:4spectra} shows a stacked spectrum of all four transits. To stack these spectra we first added the time series flux in each channel centered around the ephemeris for each observation. We then fit our transit model to the combined time series flux in each channel and binned as before (R=17) to produce the combined spectrum. The most significant features are still present at 4.5 and 5.1\,$\mu$m even with reduced overall white-light transit depth. 

    We applied the same ``look elsewhere" technique described in Section \ref{sec:app_lightcurves} and shown in Figure \ref{fig:lookelsewhere} to test the depth measurements in each spectral channel. Similar to the results in combined-light, we find that the standard deviation in measured depths is 1.5$\times$ the uncertainty we are reporting. The significance of the feature in the 5.1\,$\mu$m bin as shown in the JWST spectrum in Figure \ref{spectrum} is 6.8$\sigma$, or 4.5$\sigma$ with a 1.5$\times$ increase in the uncertainties.

\bigskip
\bigskip
\section{Optical data references} \label{sec:app_opt_data}

\vspace{-0.25cm}

\begin{table}[hbt!]
\renewcommand{\arraystretch}{1.2}
\caption{Dust optical data sources. For details on the wavelength ranges please refer to Table 1 in \citet{Kitzmann2018}.}      
\label{tab:dust}            
\centering                  
\begin{tabular}{l l}        
\hline\hline                
\noalign{\smallskip}
Condensate & Reference \\   
\hline                      
\noalign{\smallskip}
\ce{Al2O3}\,[s]&  \citet{AlO3_Begemann}$^{*t}$; \citet{Al2O3_Koike}$^{t}$ \\
CaTiO$_3$\,[s] & \citet{CaTiO3_Posch}$^{t}$; \citet{CaTiO3_Ueda}$^{f}$\\
Fe\,[s]&  Lynch \& Hunter in \citet{Palik1991}$^{t}$ \\ 
Fe$_{0.4}$Mg$_{0.6}$O\,[s]& \cite{FeO_Henning}$^{*t}$\\
Fe$_{0.7}$Mg$_{0.3}$O\,[s]& \cite{FeO_Henning}$^{*t}$\\
\ce{Fe2SiO4}\,[s]& unpublished$^{*}$\\
FeO\,[s]&  \cite{FeO_Henning}$^{*t}$ \\ 
FeS\,[s]&  \cite{FeS_Pollack}$^{t}$; \cite{FeS_Henning}$^{t}$ \\
Mg$_2$SiO$_4$\,[s]&  \cite{Mg2SiO4_Jager}$^{*t}$ \\ 
MgSiO$_3$\,[s]&  \cite{Mg2SiO4_Jager}$^{*t}$ \\ 
\ce{MgAl2O4}\,[s]& \cite{Zeidler2011}; Tropf \& Thomas in \cite{Palik1991}$^{t}$\\
SiC\,[s] & \cite{Laor1993}$^{t}$\\
SiO$_2~$\,[s] & \citet{FeS_Henning}$^{*t}$; Philipp in \citet{Palik1985}$^{t}$\\
\noalign{\smallskip}
\hline                      
\multicolumn{2}{l}{\small We note we use the amorphous (sol-gel) data for Mg$_2$SiO$_4$[s] and MgSiO$_3$[s], and the amorphous data for SiO$_2$[s].}\\
\multicolumn{2}{l}{\small *Data from the Database of Optical Constants for Cosmic Dust, Laboratory Astrophysics Group of the AIU Jena.}\\
\multicolumn{2}{l}{\small $^t$Data from a printed or digital table.}\\
\multicolumn{2}{l}{\small $^f$Data from a figure.}\\

\end{tabular}
\end{table}

\end{document}